\documentclass[12pt, reqno, oneside,amsart]{article}
\usepackage[
left=1.2in,
right=1.2in,
top=1.in,
bottom=1.in,
includeheadfoot,
footskip=0.7in
						]{geometry} 
\usepackage{amsmath}

\usepackage{cite}
\usepackage{graphicx}
\usepackage{amsfonts}
\usepackage{amssymb}
\usepackage{dsfont}		

\input epsf

\usepackage[shortlabels]{enumitem}
\usepackage{braket}
\usepackage{lipsum}
\usepackage[titletoc]{appendix}
\usepackage{enumitem}
\usepackage[dvipsnames]{xcolor}

\usepackage{hyperref}
\hypersetup{
    colorlinks=true,
    anchorcolor=red,
    urlcolor=magenta,
    citecolor=red,
    linkcolor=cyan}

\usepackage{empheq}

\usepackage{mathrsfs}

\usepackage{stmaryrd}

\newcommand{\upmapsto}{\rotatebox[origin=c]{90}{$\mapsto$}\mkern2mu}

\usepackage{subfigure}

\usepackage{amsmath} 

\usepackage{slashed}

\usepackage{tikz-cd}

\numberwithin{equation}{section}

\usepackage{etoolbox}
\usepackage{titlesec}
\usepackage{titletoc}

\titleformat{\section} 
	{
	\normalsize
	\large
	\bfseries
	} 
{\thesection.}       
{0.3em}                  
{}

\titlespacing{\section}
{0pt}                  
{20pt}                  
{13pt}                  

\titleformat{\subsection} 
[hang]                 
{
\bfseries
} 
{\thesubsection.}       
{0.3em}                  
{}                     

\titlespacing{\subsection}
{0pt}                  
{20pt}                  
{15pt}                  

\titleformat{\subsubsection} 
[hang]
{
\normalsize
\bfseries
} 
{\thesubsubsection.}    
{0.3em}                
{}                     

\titlespacing{\subsubsection}
{0pt}                  
{15pt}		        
{10pt}                  

\def\a{\alpha}
\def\b{\beta}
\def\s{\sigma}

\def\la{\lambda}

\def\ga{\gamma}
\def\Ga{\Gamma}

\def\e{\epsilon}

\def\fra{{\frak a}}
\def\frb{{\frak b}}
\def\frc{{\frak c}}

\def\frh{{\frak h}}

\def\frp{{\frak p}}

\def\frr{{\frak r}}
\def\frs{{\frak s}}
\def\frt{{\frak t}}

\def\rm{\mathrm}
\def\cal{\mathcal}
\def\scr{\mathscr}
\def\bs{\boldsymbol}
\def\bb{\mathbb}

\def\pa{\partial}

\def\be{\begin{equation}}
\def\ee{\end{equation}}
\def\br{\begin{eqnarray}}

\def\er{\end{eqnarray}}
\def\bsub{\begin{subequations}}
\def\esub{\end{subequations}}

\def\NS{\mathrm{NS}}
\def\R{\mathrm{R}}

\def\Oint{O^{(\rm{int})}_{[2]}}

\def\Oincov{\hat O^{(\rm{int})}_2}

\def\lord{\big(}
\def\rord{\big)}

\def\:{{\Bigg(}}

\def\bbS{{\mathbb S}}

\def\Scov{\hat\Sigma_\rm{cover}}
\def\Sbase{{\mathbb S}_\rm{base}}

\def\Bell{Y}

\def\gaO{\gamma_0}

\def\gaInf{\gamma_\infty}

\title{Four-point functions with fractional R-symmetry excitations in the D1-D5 CFT}

\date{} 

\usepackage{authblk}

\author[1]{\normalsize V.~A.~Souza Alves\thanks{victor.aalves@ufpe.br}}
\author[1]{\normalsize Andre Alves Lima\thanks{alves.lima@ufpe.br}}
\author[2]{\normalsize G.~M. Sotkov\thanks{galen.sotkov@ufes.br}}
\author[3]{\normalsize M. Stanishkov\thanks{marian@inrne.bas.bg}}

\affil[1]{\textit{\footnotesize Departamento de Física, Universidade Federal de Pernambuco\\ 50670-90, Recife, Brazil}}
\affil[2]{\textit{\footnotesize Departamento de Física, Universidade Federal do Esp\'irito Santo, 29075-900, Vit\'oria, Brazil}}
\affil[3]{\textit{\footnotesize Institute for Nuclear Research and Nuclear Energy, Bulgarian Academy of Sciences, 1784 Sofia, Bulgaria}}

\begin{document}

\begin{titlepage}

\maketitle

\vspace{-.6cm}

\begin{center}
\itshape{In memory of Ivan Todorov}
\end{center}

\vspace{.1cm}

\begin{abstract}

We study correlation functions with fractional-mode excitations of the R-symmetry currents in D1-D5 CFT. We show how fractional-mode excitations lift to the covering surface associated with correlation functions as a specific sum of integer-mode excitations, with coefficients that can be determined exactly from the covering map in terms of Bell polynomials. We consider the four-point functions  of fractional excitations of two chiral/anti-chiral NS fields, Ramond ground states and the twist-two scalar modulus deformation operator that drives the CFT away from the free point. We derive explicit formulas for classes of these functions with twist structures $(n)$-$(2)$-$(2)$-$(n)$ and $(n_1)(n_2)$-$(2)$-$(2)$-$(n_1)(n_2)$, the latter involving double-cycle fields. The final answer for the four-point functions always depends only on the lift of the base-space cross-ratio. We discuss how this relates to Hurwitz blocks associated with different conjugacy classes of permutations, the corresponding OPE channels and fusion rules.

{\footnotesize 
\bigskip
\noindent
\textbf{Keywords:}

\noindent
Symmetric orbifold CFT; D1-D5 CFT; Correlation functions.
}

\end{abstract}

\pagenumbering{gobble}

\end{titlepage}

\pagenumbering{arabic}

\tableofcontents

\section{Introduction and summary} \label{sec:Introduction}%

The `D1-D5 CFT' is the two-dimensional ${\cal N}=(4,4)$ superconformal field theory on the symmetric orbifold $({\mathbb T}^4)^N/S_N$ holographicaly dual to a system of D1- and D5-branes which, in the large-$N$ limit provides  important examples of black holes in string theory \cite{Strominger:1996sh,Breckenridge:1996is,Maldacena:1998bw,Maldacena:1997re,David:2002wn}. 
Although the CFT dual to the supergravity solutions is strongly coupled, its moduli space has a point where the fields are free. This free orbifold is the exact dual of the superstring on ${\rm{AdS}}_3 \times \bbS^3 \times {\mathbb T}^4$ with one unit of NS-NS flux  \cite{Dabholkar:2007ey,Gaberdiel:2007vu,Dei:2019osr,Eberhardt:2018ouy,Eberhardt:2019qcl,Eberhardt:2019ywk,Gaberdiel:2020ycd}.

At the free orbifold point, the D1-D5 CFT is a non-trivial  theory \cite{Lunin:2001pw,Lunin:2000yv}.%
	\footnote{%
	The free orbifold CFT has been largely studied; some important references related to the aspects considered in the present paper are
	\cite{
	deBeer:2019ioe,
	Burrington:2012yn,
	Burrington:2018upk,
	Burrington:2022rtr,Burrington:2022dii,Burrington:2015mfa,
	Pakman:2009ab,Pakman:2009zz,Pakman:2009mi,
	Roumpedakis:2018tdb,
	Dei:2019iym}
	}
Fields/states can be classified by their  R-charges $q$, their conformal dimensions $h$ and by a permutation $g \in  S_N$. A single cycle of length $n$ defines a $n$-twisted sector of the Hilbert space (often called a ``single particle'', or ``string component''). In the Neveu-Schwarz (NS) sector, there is a collection of chiral primary states $\ket{\s^{\fra \dot \fra}}_n$, where $\fra, \dot\fra$ are a collection of $SU(2)$ indices,
which satisfy the BPS bound: $q^\fra_n = h^\fra_n$ and $\bar q^{\dot \fra}_n = \bar h^{\dot\fra}_n$. These states are called ``1/4-BPS'' (they preserve 8 supercharges). In the full Hilbert space we have tensor products corresponding to the factorization of permutations into disjoint cycles,
\be	\label{14BPS}
\Big| \text{1/4-BPS} \Big\rangle
=
\prod_{i} \prod_{n=1}^N \big( \ket{\s^{\fra_i \dot \fra_i}}_n \big)^{N_n^i} ,
\qquad
\sum_{i,n} n N^i_n = N .
\ee

In this paper we will be interested in fractional excitations of 1/4-BPS states, that is, states produced by the action of fractional modes on (\ref{14BPS}).
Specifically, we are interested in modes of the R-current $J^a$. The component $J^3$ defines the R-charge as the eigenvalue of $J^3_0$, while the components $J^\pm$ behave as ``raising and lowering operators'' in $\frak{su}(2)$.
The Virasoro and R-current modes form a subalgebra of the superalgebra, and their fractional modes appear in OPEs of twisted fields.  For example, it was shown in \cite{deBeer:2019ioe} that the fusion of the NS chiral primaries $\s^{-}_{[n]}$ with $q^-_n = h^-_n = \frac12 (n-1)$
contains fractional Virasoro and R-current modes,
\be	\label{OPEsigmmInt}
\big[ \s^{-}_{[2]} \big] \times \big[ \s^{-}_{[n-1]} \big]
	=
	\big[ \s^{-}_{[n]} \big]
	+
	\big[ J^3_{- \frac{2}{n}} \s^{-}_{[n]} \big] 
	+
	\big[ J^3_{- \frac{3}{n}} \s^{-}_{[n]} \big] 
	+ 
	\big[ L_{- \frac{2}{n}} \s^{-}_{[n]} \big]
	+ 
	\big[ L_{- \frac{3}{n}} \s^{-}_{[n]} \big]
	+
	\cdots
\ee
These fractional excitations are therefore an important part of the dynamical data of the orbifold CFT.
There is evidence of a universal structure of the OPE algebra of bare twists and their fractional Virasoro descendants 
\cite{Burrington:2018upk,Burrington:2022dii,Burrington:2022rtr}.
There is also  indication of a similar structure for NS chiral primaries related to bare twists by spectral flow of the ${\cal N}=4$ superalgebra \cite{deBeer:2019ioe}, which include the case in (\ref{OPEsigmmInt}). 

An important part of the observables of a CFT are the correlation functions. 
For twisted fields, the natural way to describe them was developed by Lunin and Mathur \cite{Lunin:2000yv}. One constructs ramified coverings of the sphere, the ramification structure determined by the twists such as to trivialize the monodromies.
A twisted field ${\scr O}_{[n]}$ on the ``base sphere'', $\Sbase$, then lifts to the covering surface $\Scov$ as an untwisted field $\hat {\scr O}_n$ inserted at a ramification point. 
We will show that fractional excitations $J^a_{k/n} {\scr O}_{[n]}$ lift as a superposition of integer-mode excitations $J^a_{\ell} \hat {\scr O}_n$, with coefficients that are fully computable from the covering map. 

We will consider in detail the excitations of three types of fields: NS chirals $\s^\fra_{[n]}$, anti-chirals $\s^{\fra \dagger}_{[n]}$ and some ``multiplet fields'' $\s^{\fra (r)}_{[n]}$ defined by Lunin and Mathur in \cite{Lunin:2001pw}, constructed by applying the operator $(J^-_0)^r$ to NS chiral states so as to decrease the R-charge without changing the dimension.
We find that these fields do \emph{not} lift to Kac-Moody primaries on the covering.
For instance, the NS chiral $\s^\fra_{[n]}$ lifts to $\Scov$ as a field $\hat \s^\fra_n$ for which
\be	\label{INTROJminhasas}
J^-_\ell \hat \s^\fra_n \neq 0 
\quad \text{for a set of $\ell > 0$} .
\ee
The same happens for $J^+_\ell \hat \s^{\fra \dagger}_n$, and for excitations of $\hat\s^{\fra(r)}_n$ by any $J^a$ component.
As we also find, (\ref{INTROJminhasas}) implies that, on the base sphere,
\be	\label{INTROJminhasasBas}
 J^-_{k/n} \s^\fra_{[n]} \neq 0 
\quad \text{for $0 < k < n-1$} .
\ee

We have said that fractional modes lift to a superposition of integer modes.
Correlation functions with fractional-mode excitations on $\Sbase$ therefore lift to superpositions of functions with integer-mode excitations. 
On $\Scov$ there is only one copy of the $\frak{su}(2)_1$ current algebra.
Assuming $\Scov = \bbS_\rm{cov}$ has the topology of a sphere, we can use ordinary Ward identities to try and reduce each function with integer-mode excitations to correlators \emph{without} any excitations.
In \cite{Burrington:2022dii} this is shown to be possible for functions with fractional excitations by Virasoro modes, as long as the fields being excited lift to primaries on the covering. 
Here we find that, for excitations by $J^a$, the non-vanishing of modes like (\ref{INTROJminhasas}) and (\ref{INTROJminhasasBas}) \emph{prevent} a similar reduction, in general. 
For instance, the function
\be	\label{IntoJpmNS4pt}
\Big\langle
\Big( J^{+}_{\frac{k'}{n'}} \s^{\fra \dagger}_{[n']}\Big) (\infty)
\; \s^{\frb \dagger}_{[n_1]}(z_1)
\; \s^{\frb}_{[n_2]}(z_2)
\Big( J^-_{-\frac{k}{n}} \s^{\fra}_{[n]}\Big) (0)
\Big\rangle 
\ee
cannot be reduced to a correlator containing only NS chirals and anti-chirals, while
\be	\label{IntoJ3NS4pt}
\Big\langle
\lord J^{3}_{\frac{k'}{n'}} \s^{\fra \dagger}_{[n']} \rord (\infty)
\; \s^{\frb \dagger}_{[n_1]}(z_1)
\; \s^{\frb}_{[n_2]}(z_2)
\lord J^3_{-\frac{k}{n}} \s^{\fra}_{[n]} \rord (0)
\Big\rangle 
\ee
can, as we show by explicit computation.

An important part of the problem of computing explicit correlation functions in the orbifold is that we need the covering map fixed by the corresponding twist structure. Here we take as a working case an important class of four-point functions with twist structure $(n)$-$(2)$-$(2)$-$(n)$, i.e.~four single-cycle fields, two of them with twist of length 2 and two with length $n$.
This twist structure is, in itself, conveniently simple while not trivial, and the covering map with the topology of a sphere is well-known. 
Most importantly, this is the structure of a class of functions with a deformation operator that is very important in perturbation theory away from the free orbifold point.

There are twenty supersymmetry-preserving marginal deformations of the free orbifold theory that move it towards a strongly-coupled region of moduli space
\cite{David:2002wn,
Larsen:1999uk,
Gava:2002xb,
Avery:2010qw}. 
Most are built from the bosons $\pa X$ and are untwisted. The others are excitations of the NS chiral field $\s^{+ \dot +}_{[2]}$ by the supercharge; we will work here with the only operator of this latter type which is a singlet under the $SO(4)$ associated with ${\bb T}^4$. We call it $\Oint$.
It defines the perturbed action
\be	\label{def_cft}
S_\rm{orb} + \lambda \int \!\! d^2z \ \Oint (z,\bar z) .
\ee
From the point of view of the fuzzball/microstate program
\cite{Mathur:2024ify,Skenderis:2008qn,Bena:2015bea} 
(see 
\cite{Bena:2022ldq,Bena:2022rna,Shigemori:2020yuo} 
for reviews), understanding the effect of the deformation is important because the CFT dual to the growing number of supergravity solutions that have been constructed over the last years
is strongly coupled, i.e.~it is a theory with large $\la$. The free orbifold is used as a computational tool which must be complemented by information about how and which states ``lift'', i.e.~are renormalized under the deformation
\cite{Hughes:2023apl,Hughes:2023fot,Benjamin:2021zkn,Guo:2020gxm,Guo:2019ady,Guo:2019pzk,Hampton:2018ygz,Keller:2019suk}.
From the point of view of strings on ${\rm{AdS}}_3 \times \bbS^3 \times {\mathbb T}^4$, the deformation (\ref{def_cft}) corresponds to turning on a Ramond-Ramond flux, which leads to the problem of integrability for AdS$_3$ backgrounds (see e.g.~\cite{Seibold:2024qkh,Demulder:2023bux}
for reviews). This is an important problem with important recent developments 
\cite{Gaberdiel:2023lco,Frolov:2023pjw}, in the context of which the perturbative study of renormalized anomalous dimensions is also relevant 
\cite{Gaberdiel:2024nge,Gaberdiel:2024dfw,Fabri:2025rok}.

With this in mind, here we compute the functions
\be	\label{INTROJandOint}
\Big\langle
\Big( J^{a \dagger}_{\frac{k}{n}} \s^{\fra \dagger}_{[n]} \Big) (\infty) \
\Oint(1) \
\Oint(u) \
\Big(J^a_{-\frac{k}{n}} \s^{\fra}_{[n]}\Big)  (0)
\Big\rangle .
\ee
They can be used to compute the anomalous dimensions of the field $J^a_{-k/n} \s^\fra_{[n]}$ at second order in perturbation theory.
As expected, we show that the functions with $J^3$ excitations can be reduced to correlators without excitations, much like what happens with (\ref{IntoJ3NS4pt}). For $J^\pm$ excitations, the conditions (\ref{INTROJminhasas}) and (\ref{INTROJminhasasBas}) would make us expect that the reduction is not possible but, in this particular case, there is a simplification and it turns out that they, as well, can be reduced.
That is, all functions like (\ref{INTROJandOint})  lift to a superposition of 
\be	\label{INTROJandOintCOV}
\begin{aligned}
&
\Big\langle
\Big( J^{a\dagger}_{\ell'} \hat \s^{\fra \dagger}_{n}\Big) (\infty) 
\Oincov(t_1) 
\Oincov(x) \;
\Big( J^a_{- \ell} \hat \s^\fra_n\Big) (0)
\Big\rangle 
\\
&\qquad
	=
	{\scr D}^{\{a;k;\fra\}}_{\ell',\ell}(x) 
	\Big\langle
	\hat \s^{\fra \dagger}_{n} (\infty) 
	\Oincov(t_1) 
	\Oincov(x) \;
	\hat \s^\fra_n (0)
	\Big\rangle .
\end{aligned}
\ee
The correlators in the RHS have been computed in the literature, hence our determination, here, of the functions ${\scr D}^{\{a;k;\fra\}}_{\ell',\ell}(x)$ completes the computation of (\ref{INTROJandOint}).

It was shown in \cite{Lima:2022cnq} that, under certain assumptions, at large $N$, the leading contribution to functions similar to (\ref{INTROJandOint}) but with generic twists at $z=0,\infty$ comes, after some factorizations, from functions with \emph{double}-cycle fields. Thus we consider here correlators of the form
\be	\label{INTROdoublCycl0}
	\Big\langle
	\Big[ 
	\Big( J^{a_1\dagger}_{\frac{k_1}{n_1}} \s^{\frak a_1\dagger}_{[n_1]} \Big)  \
	\Big( J^{a_2\dagger}_{\frac{k_2}{n_2}} \s^{\frak a_2\dagger}_{[n_2]} \Big) 
	\Big] (\infty) \
	\Oint(1) \
	\Oint(u) \
	\Big[ 
	\Big( J^{a_1}_{-\frac{k_1}{n_1}} \s^{\frak a_1}_{[n_1]}\Big)  \
	\Big( J^{a_2}_{-\frac{k_2}{n_2}} \s^{\frak a_2}_{[n_2]} \Big)
	\Big] (0)
	\Big\rangle .
\ee
The covering map for these functions is also known, and they lend themselves to the same methods that we use for single-cyle fields.
Again, if both ``string components'' are excited by $J^3$ we show, by explicit computation, that the function can be reduced to a correlator without excitations as in (\ref{INTROJandOintCOV}), with a more complicated coefficient function
${\scr D}^{\{3,3;k_1, k_2;\fra_1,\fra_2\}}_{\ell'_1,\ell'_2,\ell_1,\ell_2}(x)$.
If we have $J^\pm$ excitations, however, this is not true: the simplifications that lead to (\ref{INTROJandOintCOV}) in this case do not help us anymore, and Eqs.(\ref{INTROJminhasas}) and (\ref{INTROJminhasasBas}) prevent us from simplifying the function, as it was the case with (\ref{IntoJpmNS4pt}).
Anyhow, our computation of ${\scr D}^{\{3,3;k_1, k_2;\fra_1,\fra_2\}}_{\ell'_1,\ell'_2,\ell_1,\ell_2}$, combined with the functions without excitations that can be found in \cite{Lima:2022cnq}, complete the computation of (\ref{INTROdoublCycl0}) for the $J^3$ case.
This can also be used to determine anomalous dimensions of double- (or multi-) cycle fields at second order in perturbation theory.

Working on similar problems, the authors of \cite{Burrington:2022rtr,Burrington:2022dii}, have found that the only information about the covering map that survives in the final answer for the four-point functions is one lone parameter that describes both the covering and the base spaces cross-ratios. We find the same thing here, as it was the case also for the large classes of functions of (unexcited) primaries computed, say, in \cite{Lima:2022cnq}.
In our notations for the correlators (\ref{INTROJandOint}) and (\ref{INTROJandOintCOV}), the parameter in question is $x \in \Scov$, the lift on the covering surface of the base-space cross-ratio $u \in \Sbase$. All the lifted correlation functions (even those which cannot be reduced to correlators without excitations) are, in the end, functions of $x$ exclusively (with $n$, $k$, etc.~as parameters). We give several concrete examples of this. The relation between $x$ and $u$, hence the relation between the covering-surface and the base-space functions, is interesting. As shown in \cite{Pakman:2009ab,Pakman:2009zz,Pakman:2009mi}, it can be formulated in terms of Hurwitz theory. We use this language to give here an analysis of OPE channels inside four-point functions like (\ref{INTROJandOint}), similar to what was done in \cite{Lima:2022cnq}. This allows us to study some fusion rules involving fractional-mode excited fields.
As an example, we use some of our four-point functions to confirm the existence of the $J^3$-excited fields (which are actually Virasoro primary) in the fusion rule (\ref{OPEsigmmInt}), as predicted in \cite{deBeer:2019ioe}.

Finally, we consider Ramond fields, i.e.~the fields that produce the collection of Ramond ground states in the $n$-twisted sector. 
They are related to NS chiral fields by spectral flow of the superalgebra.
Correlators involving Ramond ground states have a caveat concerning the fact that the vacuum is not in the Ramond sector, so the untwisted copies of the orbifold are not trivial. The combination of permutations in a twisted correlation function then leads, for example, in four-point functions lifting to (a combination of) six-point functions on the covering. 
In spite of this complication, it turns out that the lifted Ramond ground states are much more natural objects from the point of view of the action of the R-currents: they are the ones that transform as Kac-Moody primaries on the covering. This allows us to reduce functions containing only Ramond fields, their excitations, and/or $\Oint$ to a combination of correlators without any excitations, ``rotated'' by $\frak{su}(2)$ representation matrices.

\section{Fractional excitations in the D1-D5 CFT}

The basic definitions for the D1-D5 CFT at the free orbifold point are given in Appendix \ref{SectNotationsD1D5}; we use the same conventions as \cite{Avery:2010qw}.
In this section, which also serves to define notations, we will focus on the most important definitions used in the development of the paper.

\subsection{Fractional modes}

The generators of the ${\cal N} = (4,4)$ superalgebra of the ``seed CFT'' (the $N$ copies of which are identified under $S_N$ in the orbifold)
are the stress tensor $T(z)$, the supercurrents $G^{\a A}(z)$ and the R-currents $J^a(z)$, along with the corresponding antiholomorphic fields; see (\ref{TJGasXpsi}).
The current algebra can be found e.g.~in \cite{Avery:2010qw}.
Here we will be mostly interested in the subalgebra
\be	\label{JaJbOPEdef}
J^a(z) J^b(z') 
	\sim 
	\frac{c}{12} \frac{\delta^{ab}}{(z-z')^2}
	+ \frac{i \e^{abc} J^c(z')}{z-z'}
\ee
where $\e^{abc}$ is the anti-symmetric symbol with $\e^{123} = 1$ and $c = 6$ is the central charge of the seed CFT.
These OPEs are equivalent to the Kac-Moody $\frak{su}(2)_1$ algebra for the modes,%
	\footnote{
	See \cite{Todorov:1985ap,Kac:1985wdn,Furlan:1989vv}.
	The $n$-twisted sector can be regarded as an orbifold by ${\bb Z}_n$, whose algebras have been considered in \cite{Kac:1996nq}.
	}
\be	\label{JaJbmodealge}
[J^a_m ,J^b_n] =  i \e^{abc} J^c_{m+n} + \frac{c}{12} \, m \, \delta_{m+n,0} \, \delta^{ab}.
\ee

In the full orbifold, we have $N$ copies of these currents labeled by an index $I = 1,\cdots, N$. 
The ground state of the $n$-twisted sector of Hilbert space, $\ket{\s}_{n}$, corresponds to the insertion of a `bare twist' operator $\s_{(n)}$, with the cyclic permutation $(n) = (1,\cdots, n) \in S_N$.
The operators
\be	\label{DiagonJsum}
\begin{aligned}
J^a_{\{k\}}(z) &\equiv \sum_{I=1}^n \exp \left( - \frac{ 2\pi i k(I-1)}{n} \right) J^a_I (z) ,
\qquad	k \in {\bb Z} ,
\end{aligned}
\ee
diagonalize the action of the permutation.
The combination is such that, in the presence of the twist insertion, the monodromy is diagonal with eigenvalues $e^{- 2\pi i k / n}$,
\be	\label{MonoDiagCurr}
\begin{aligned}
J^a_{\{k\}}( e^{2\pi i} z) \s_{(n)}(0) &= e^{-2\pi i k / n} J^a_{\{k\}}(z)\s_{(n)}(0) .
\end{aligned}
\ee
Fractional modes of $J^a$ can then be defined as the field 
\be	\label{FracModsJA}
 J^a_{- \frac{k}{n}} \s_{(n)}(z_*) 
	\equiv	
	\oint_{z_*} \frac{d z}{2\pi i}
	(z-z_*)^{- \frac{k}{n}} 
	J^a_{\{k\}} (z) \s_{(n)}(z_*).
\ee
They satisfy a closed algebra with the fractional Virasoro modes $L_{k/n}$ \cite{deBeer:2019ioe,Roumpedakis:2018tdb}
\bsub	\label{ModeAlgebra}
\be
\begin{aligned}
\left[ J^a_{\frac{k}{n}} , J^b_{\frac{k'}{n}} \right]
	&=
	i \e^{abc} J^c_{\frac{k+k'}{n}} + \frac{k}{2}  \, \delta_{k+k',0} \delta_{a,b} ,
\\
\left[ L_{\frac{k}{n}} , J^a_{\frac{k'}{n}} \right]
	&=
	- \frac{k'}{n} J^a_{\frac{k+k'}{n}} ,
\\
\left[ L_{\frac{k}{n}} , L_{\frac{k'}{n}} \right]
	&=
	\frac{k-k'}{n} L_{\frac{k+k'}{n}} + \frac16 \Big( \frac{k^2}{n^2} - 1 \Big) \delta_{k+k',0} ,
\end{aligned}
\ee
Instead of the Cartesian components of $J^a$, it is more convenient to use the raising/lowering basis with $J^\pm_{k/n} = J^1_{k/n} \pm i J^2_{k/n}$ and algebra
\be
\Big[ J^3_{\frac{k}{n}} , J^\pm_{\frac{k'}{n}} \Big] = \pm J^\pm_{\frac{k+k'}{n}} ,
\quad
\Big[ J^+_{\frac{k}{n}} , J^-_{\frac{k'}{n}} \Big] = 2 J^3_{\frac{k+k'}{n}} + k \delta_{k+k',0} ,
\quad
\Big[ J^3_{\frac{k}{n}} , J^3_{\frac{k'}{n}} \Big] = \frac{k}{2} \delta_{k+k',0} .
\ee
\esub

The highest-weight states in the representations of the algebra (\ref{ModeAlgebra}) are states of the type $\ket{h,q}_n$ such that, in the holomorphic sector,
\be	\label{PrimLJ}
\begin{aligned}
L_0 \ket{h,q}_n &= h \ket{h,q}_n ,
\\
J^3_0 \ket{h,q}_n &= q \ket{h,q}_n ,
\end{aligned}
\qquad
\begin{aligned}
L_\ell \ket{h,q}_n &= 0 ,
\\
J^a_\ell \ket{h,q}_n &= 0 ,
\end{aligned}
\qquad
\ell > 0 .
\ee
The BPS unitarity bound imposes $h \geq |q|$.
Fractional excitations of $\ket{h,q}_n$ by negative modes $J^a_{- k/n}$ have well-defined dimensions and charges, which can be computed from the algebra:
\be	\label{DimCharket}
\begin{aligned}
L_0 J^a_{-\frac{k}{n}} \ket{h,q}_n 
	&= (h + k/n) J^a_{-\frac{k}{n}} \ket{h,q}_n ,
\\
J^3_0 J^\pm_{-\frac{k}{n}} \ket{h,q}_n 
	&= (q\pm1) J^\pm_{-\frac{k}{n}} \ket{h,q}_n ,
\\
J^3_0 J^3_{-\frac{k}{n}} \ket{h,q}_n 
	&= q J^3_{-\frac{k}{n}} \ket{h,q}_n .
\end{aligned}
\ee

\subsection{NS chirals and multiplets}

We will be concerned with the family of ``1/4-BPS'' states in the $n$-twisted sector. We follow the notations of \cite{Shigemori:2020yuo} and define the states
\be	\label{NSchiralStates}
\begin{aligned}
&\ket{\s^{\a \dot \a}}_n
	&&\qquad \text{with} \quad
	h^\a_n = q^\a_n = \tfrac{n+\a}{2} ,
	&&\quad
	\tilde h^{\dot\a}_n = \tilde q^{\dot\a}_n = \tfrac{n+\dot\a}{2} 
\\
&\ket{\s^{\dot A \dot B}}_n,
	&&\qquad \text{with} \quad
	h^{\dot A}_n = q^{\dot A}_n = \tfrac{n}{2} ,
	&&\quad
	\tilde h^{\dot B}_n = \tilde q^{\dot B}_n = \tfrac{n}{2} 
\end{aligned}
\ee
(In the formulas for the dimensions the SU(2) indices $\a,\dot \a = \pm$ should be interpreted as $\pm 1$.)
The holomorphic parts have only one index, $\ket{\s^\fra}_n$, and we will very often use gothic lower-case for all-purpose indices, $\fra = \a, \dot A$, etc.

The states in (\ref{NSchiralStates}) are of the form $\ket{h,h}$, i.e.~are BPS, $q = h$.
Besides (\ref{PrimLJ}), it holds that $G^{+A}_{- 1/2} \ket{h,h} = 0$, so these states and the corresponding fields are ``NS chiral.''
From (\ref{DimCharket}), we see that 
\be	\label{Jplusketis0}
J^+_{-\frac{k}{n}} \ket{h,h}_n = 0 ,
\qquad k < n ,
\ee
because otherwise this state would violate the unitarity bound. It follows that we can only excite the NS chirals by fractional modes of $J^3$ or of the ``charge lowering operator'' $J^-$.
The NS chiral primary states (\ref{NSchiralStates}) themselves can be constructed by exciting the bare-twist vacuum with negative fractional modes of $J^-$ \cite{Lunin:2001pw}.
In general, for fractional excitations of a state $\ket{h,q}_n$ the algebra (\ref{ModeAlgebra}) gives
\be	\label{FracModAlgLJnotprim}
L_\ell J^a_{-\frac{k}{n}} \ket{h,q}_n
	= \frac{k}{n} J^a_{\ell -\frac{k}{n}} \ket{h,q}_n
\ee
The excited field is primary if the RHS vanishes for all $\ell > 0$. 
For the NS chiral states, we can show that, if $0 \leq k /n < 1$ the RHS  does vanishes  for $a= 3, +$, but it does not vanish  when $a= -$.

We will denote the the NS chiral fields associated with the states (\ref{NSchiralStates}) by
\be	\label{NSchiralss}
\s^{\a \dot \a}_{(n)}(z,\bar z) = \s^\a_{(n)}(z) \tilde\s^{\dot \a}_{(n)}(\bar z) ,
\qquad
\s^{\dot A \dot B}_{(n)} (z,\bar z) = \s^{\dot A}_{(n)}(z) \tilde \s^{\dot B}_{(n)}(\bar z) .
\ee
with $\ket{\s^{\fra \dot \fra}}_n \cong \s^{\fra \dot \fra}_{(n)}$.
There are also states/fields with ``mixed'' holomorphic/anti-holomorphic parts such as 
$\ket{\a \dot A}_n$ with $h = q = \frac12(n+\a)$ and $\tilde h = \tilde q = \frac12 n$, but we will not consider them in our examples.
We will often work only with the holomorphic parts $\s^\fra_{(n)}(z)$.
The anti-chiral operators,
$\s^{\fra\dagger}_{(n)}$,
have opposite R-charges and the same dimensions, i.e.~$\ket{\s^{\fra \dagger}}_n = \ket{h^\fra_n , - h^\fra_n}_n$.

The fields above are fermionic excitations of the bare twist $\s_{(n)}$ and will be more conveniently expressed in bosonized language (\ref{FermionsBoson}). 
In the free CFT, we can express all of them as exponentials of a pair (in each copy $I$) of bosons $\phi_{1,I}$ and $\phi_{2,I}$
The holomorphic part of the NS chirals read
\bsub \label{0022NSchi}
\begin{align}
\s^{\pm}_{(n)} (z) 
	&=
	e^{
	 i \frac{n\pm1}{2n} \sum_{I = 1}^n
	 ( \phi_{1,I} - \phi_{2,I} )
	} 
	\s_{(n)}(z)
\\
\s^{\dot 1}_{(n)}(z)
	&=
	e^{
	 i \sum_{I = 1}^n
	 \left( \frac{n-1}{2n}  \phi_{1,I} - \frac{n+1}{2n}  \phi_{2,I} \right)
	} 
	\s_{(n)}(z)
\\
\s^{\dot 2}_{(n)}(z)
	&=
	e^{
	 i \sum_{I = 1}^n
	 \left( \frac{n+1}{2n}  \phi_{1,I} - \frac{n-1}{2n}  \phi_{2,I} \right)
	} 
	\s_{(n)}(z)
\end{align}
\esub
(Exponentials are always implicitly normal-ordered.)
There are similar formulas for the anti-holomorphic parts $\s^{\dot\fra}(\bar z)$, in which we replace $\phi_{i,I}(z)$ by $\tilde \phi_{i,I}(\bar z)$.

\bigskip

The zero modes $J^a_0$ can be used to generate $\frak{su}(2)$ multiplets without changing the conformal dimensions of states.
Starting from an NS chiral state $\ket{q^{\frak a}_n,q^{\frak a}_n}_n$, with dimension $h^{\frak a}_n = q^{\frak a}_n$, as a highest weight vector in the $\bs{(2q^{\frak a}_n +1)}$-representation of the $\frak{su}(2)$ generated by the $J^a_0$, the other states in the multiplet are constructed by lowering the charge with $J^-_0$ to obtain the vector
\be
\begin{aligned}
& 
\ket{\s^{\fra (r)}}_n 
\equiv
\ket{q^{\frak a}_n,q^{\frak a}_n-r}_n
	= 
	{\cal A}^{\fra (r)}_n
	(J^-_0)^r
	\ket{q^{\frak a}_n,q^{\frak a}_n}_n ,
\quad
	{\cal A}^{\fra (r)}_n
	=
	\sqrt{ \frac{(2q^{\frak a}_n - r)!}{r! (2q^{\frak a}_n)!} }
\end{aligned}
\ee
We are following the definitions of Lunin and Mathur in \cite{Lunin:2001pw}. 
${\cal A}^{\fra (r)}_n$ is a normalization factor, see Eq.(6.9) of \cite{Lunin:2001pw}.
In our notations, the fields associated with these states are
$\ket{q^{\frak a}_n,q^{\frak a}_n - r}_n \cong \s^{\frak a (r)}_{(n)} (z_*)$
where we define
\be	\label{qqrfield}
	\s^{\frak a (r)}_{(n)} (z)
	\equiv	
	{\cal A}^{\fra (r)}_n
	\oint_{z} \frac{dz_1}{2\pi i} J^-(z_1)
	\cdots
	\oint_{z} \frac{dz_d}{2\pi i} J^-(z_d)
	\s^{\frak a}_{(n)} (z) .
\ee
These multiplet states are Virasoro primary, as can be checked from the mode algebra: since
$[L_\ell , J^-_0] = 0$
then 
$L_\ell \s^{\frak a (r)}_{(n)} = 0$ for $\ell > 0$.

\bigskip

The most important dynamical data in a CFT are its correlation functions.
Lunin and Mathur \cite{Lunin:2001pw,Lunin:2000yv} have shown that each correlator defines a covering surface $\Scov$ with ramified points defined by the cycles of the twist insertions on the Riemann sphere base space $\Sbase$. 
On $\Scov$ the monodromies are trivialized, and there is only one copy of the seed ${\cal N} = (4,4)$ CFT.
We will take $t,\bar t$ to be complex coordinates on $\Scov$ and $z, \bar z$ coordinates on $\Sbase$, and we assume the existence of the appropriate covering map $t \mapsto z$.
Path integrals on $\Sbase$ can be determined from path integrals on $\Scov$ and vice-versa. Given a twisted field ${\scr O}_{[n]}(z)$ on $\Sbase$, we say that it ``lifts'' to $\Scov$ as a field $\hat {\scr O}_n(t)$, and write
\be
{\scr O}_{[n]}(z) \mapsfrom \hat {\scr O}_n(t).
\ee
We will use hats to indicate fields that have been lifted to the covering surface.
The covering surface is by construction \emph{untwisted}, the subscript indicates only that $n$ is a parameter. The lifting should implicitly be understood as holding inside path integrals, but it is convenient to speak of the behavior of operators individually.

The NS chirals in their exponential form (\ref{0022NSchi}) lift to 
\bsub \label{0022NSchiCOVER}
\begin{align}
\hat \s^{\pm}_{n} (t) 
	&=
	b_*^{- \frac{(n \pm 1)^2}{4 n}}
	e^{ i \frac{n\pm1}{2} (\phi_{1} - \phi_{2} )} (t)
\\
\hat \s^{\dot 1}_{n}(t)
	&=
	b_*^{- \frac{1+n^2}{4 n}}
	e^{i \left( \frac{n-1}{2}  \phi_{1} - \frac{n+1}{2}  \phi_{2} \right) } (t)
\\
\hat \s^{\dot 2}_{n}(t)
	&=
	b_*^{- \frac{1+n^2}{4 n}}
	e^{i \left( \frac{n+1}{2}  \phi_{1} - \frac{n-1}{2}  \phi_{2} \right) } (t)
\end{align}
\esub
where $b_*$ is a parameter that locally defines the covering map near the ramification point created by the twist (see Eq.(\ref{ztgenercoveC}) below). 
Anti-chiral fields lift to $\hat \s^{\frak a \dagger}_n(t)$ given by the same formulas (\ref{0022NSchiCOVER}) but with opposite sign in the exponents.
The multiplet fields (\ref{qqrfield}) lift to the covering as 
\be	\label{qqrfieldCOVER}
\hat \s^{\fra (r)}_{n} (t)
	\equiv
	{\cal A}^{\fra (r)}_n
	\oint_{t} \frac{dt_1}{2\pi i} J^-(t_1)
	\cdots
	\oint_{t} \frac{dt_d}{2\pi i} J^-(t_d)
	\hat \s^{\frak a}_{n} (t) 
\ee
where $\nu_{\frak a}$ are the exponents of $b_*$ for the different NS chirals that can be read in (\ref{0022NSchiCOVER}). 
The R-symmetry currents on the covering are very conveniently expressed in terms of the bosonized fermions,
\begin{align}
J^3(t) &= \frac{i}{2} \Big( \pa \phi_1(t) - \pa \phi_2(t) \Big) ,
		\label{bosonJ3}
\\
J^\pm (t) &=  e^{ \pm i (\phi_1 -  \phi_2)} (t) .
	\label{bosonJpm}
\end{align}
(We are ignoring cocycles; see \cite{Burrington:2012yq} and \cite{Burrington:2015mfa} for a very detailed account of bosonization in the ${\cal N} = (4,4)$ SCFT.)

\bigskip

In fact, the twisted states that we have been describing are not physical because they are not gauge invariant. The symmetric group acts on the cyclic permutation $(1,2,\cdots, n)$ by conjugation to produce every other cyclic permutation with length $n$ involving different copies. 
Thus if ${\scr O}_{(n)}$ is a twisted field with the standard cycle $(1,2,\cdots,n)$, the corresponding physical field is the sum over the entire equivalence class
\be	\label{scrOconj}
{\scr O}_{[n]} \equiv \frac{1}{{\scr S}_n} \sum_{g \in S_N} {\scr O}_{g (1,2,\cdots,n) g^{-1}} ,
\qquad
{\scr S}_n = \sqrt{N! (N-n)! n} .
\ee
The normalization factor ${\scr S}_n$ counts repeated terms in such a way that the normalizations of ${\scr O}_{[n]}$ and ${\scr O}_{(n)}$ are equal
\cite{Lunin:2001pw,Pakman:2009zz}.
The covering surface technique incorporates the sum in (\ref{scrOconj}) intrinsically: once the correlator on the covering has been computed, the map back to the physical manifold is multivalued and produces the different conjugacy classes in (\ref{scrOconj}), see \cite{Pakman:2009zz}.
This will be further discussed for our cases of interest in Sect.\ref{SectFunctsonBase}.

\section{Lifting of fractional-mode excitations}	\label{SectLiftingFracMode}

\subsection{Fractional-mode coefficients and Bell polynomials}	\label{SectAandBellPoly}

Fractional excitations 
$J^a_{k/n} \ket{{\scr O}}_n$
of a generic state $\ket{{\scr O}}_n \in {\cal H}_{(n)}$ 
correspond to the fields 
\be	\label{FracModsJ}
\lord J^a_{- \frac{k}{n}} {\scr O}_{(n)} \rord (z_*)
	\equiv	
	\oint_{z_*} \frac{d z}{2\pi i}
	(z-z_*)^{- \frac{k}{n}} 
	J^a_{\{k\}} (z) {\scr O}_{(n)}(z_*) ,
\ee
assuming that ${\scr O}_{(n)}(z)$ has been properly defined and has trivial monodromy.
This field lifts on the covering surface to
\be	\label{liftintjj}
\lord J^a_{-\frac{k}{n}} {\scr O}_{[n]} \rord (z_*)
\mapsfrom	
\oint_{t_*} \frac{d t}{2\pi i}
	[z(t)- z_* ]^{- \frac{k}{n}} 
	J^a (t) \hat {\scr O}_n(t_*) .
\ee
If $z = \infty$ is a branching point of order $n$, the modes are given by residues at infinity, which lift to
\be	\label{ModeCovJinfA}
\lord J^a_{\frac{k}{n}} {\scr O}_{[n]} \rord (\infty)
\mapsfrom
\oint_{\infty} \frac{d t}{2\pi i}
	[z(t)]^{\frac{k}{n}} 
	J^a (t) \hat {\scr O}_n(\infty) .
\ee
When there is no risk of confusion, we will sometimes omit the parentheses for economy of notation, writing, for instance, 
$ J^a_{k/n} {\scr O}_{[n]} $
instead of 
$( J^a_{k/n} {\scr O}_{[n]} )$,
etc.

The lifted field $\hat {\scr O}_n(t_*)$ has no twist, but $t_*$, defined by $z(t_*) = z_*$, is a ramification point with the appropriate order,
\begin{align}
z(t)  &= z_* + b_* (t-t_*)^n \Big[1 + c^{(*)}_{1} (t-t_*) + c^{(*)}_{2}(t-t_*)^2 + \cdots \Big]
	\label{ztgenercoveC}
\\
z(t) &= b_\infty t^n \Big[1 + c^{(\infty)}_1 t^{-1} + c^{(\infty)}_2 t^{-2} + \cdots \Big]
	\label{ztgenercoveInf}
\end{align}
where $c^{(*)}_\ell$ are coefficients. The second equation corresponds to a ramification point at $t = \infty$.
The integrands of (\ref{liftintjj}) and (\ref{ModeCovJinfA}) can then be expanded as 
\be
\begin{aligned}
[z(t)- z_* ]^{- \frac{k}{n}} 
	=
	(t- t_*)^{-k}
	\sum_{\ell=0}^\infty A_{\ell}^{(*)} (t-t_*)^\ell ,
\qquad
[z(t) ]^{\frac{k}{n}} 
	=
	t^k \sum_{\ell=0}^\infty A_{\ell}^{(\infty)} t^{-\ell} ,
\end{aligned}
\ee
where (making $\tau = 1/t$)
\begin{align}
A^{(*)}_\ell 
	&= 
	\left.
	\frac{b_*^{-\frac{k}{n}}}{\ell!} \frac{d^\ell}{dt^\ell} 
	\left[ 
	1 + c^{(*)}_{1} (t-t_*) + c^{(*)}_{2}(t-t_*)^2 + \cdots
	\right]^{- \frac{k}{n}} 
	\right|_{t=t_*} 
\label{AstaellDercc}
\\
A^{(\infty)}_\ell 
	&= 
	\left.
	\frac{b_\infty^{\frac{k}{n}}}{\ell!} \frac{d^\ell}{d\tau^\ell} 
	\left[ 
	1 + c^{(\infty)}_{1} \tau + c^{(\infty)}_{2}\tau^2 + \cdots
	\right]^{\frac{k}{n}} 
	\right|_{\tau=0}
\label{AstaellDerccInf}
\end{align}

We can extract a more explicit expression from Eq.(\ref{AstaellDercc}) if we use `Faà di Bruno's formula' (\ref{FaadiBruno}).
Looking at the RHS of Eq.(\ref{AstaellDercc}), we see that the function $g(t)$ appearing in (\ref{FaadiBruno}) is the series expansion for the covering map, and after evaluating everything at $t = t_*$, the derivatives
$g^{(n)}(t_*) = n! c^{(*)}_n$ compute the covering map coefficients $c^{(*)}_n$.
Meanwhile, for Eq.(\ref{AstaellDercc}) the function $f(x)$ appearing in (\ref{FaadiBruno}) is $f(x) = x^{-k/n}$, hence
$f^{(m)}(1) = ( - k/n )^{\underline m}$ is a \emph{falling} factorial which can be translated into
$( - k/n )^{\underline m} = 	(-)^m(k/n)_m$, 
with $(\a)_m$ the Pochhammer symbol for the \emph{rising} factorial:
\be
(k/n)_m \equiv \Ga(m + k/n) / \Ga(k/n) .
\ee
For the branch at infinity, the argument is slightly different because the power in (\ref{AstaellDerccInf}) has the opposite sign. 
In summary, we can write (\ref{AstaellDercc})-(\ref{AstaellDerccInf}) as
\be	\label{AstaellDerccBell}
\begin{aligned}
A^{(*)}_\ell 
	&= 
	\frac{b_*^{-\frac{k}{n}}}{\ell!}
	\sum_{m = 0}^\ell
	(-)^m
	(k/n)_m 
	\Bell_{\ell, m} 
		\big[ c^{(*)}_1 , 
			2! c^{(*)}_2, \cdots, 
			(\ell - m +1)! c^{(*)}_{\ell-m+1}
		\big]
\\
A^{(\infty)}_\ell 
	&= 
	\frac{b_\infty^{\frac{k}{n}}}{\ell!}
	\sum_{m = 0}^\ell
	(-)^m
	(-k/n)_m 
	\Bell_{\ell, m} 
		\big[ c^{(\infty)}_1 , 
			2! c^{(\infty)}_2, \cdots, 
			(\ell - m +1)! c^{(\infty)}_{\ell-m+1}
		\big]
\end{aligned}
\ee
where $\Bell_{\ell, m}$ are the so-called \emph{incomplete exponential Bell polynomials,} see App.\ref{AppBellPoly}.
This gives an explicit expression for the fractional-mode coefficients in terms of the expansion of the covering map.

\bigskip

Going back to Eqs.(\ref{liftintjj}) and (\ref{ModeCovJinfA}) we can now write
\begin{align}
\lord J^a_{-\frac{k}{n}} {\scr O}_{[n]} \rord (z_*)
	&\mapsfrom
	\oint_{t_*} \frac{d t}{2\pi i}
	(t- t_*)^{-k}
	\left[
	\sum_{\ell=0}^\infty A_{\ell}^{(*)} (t-t_*)^\ell 
	\right]
	J^a (t) \hat {\scr O}_n(t_*) ,
	\label{ModeCovJ}
\\
\lord J^a_{\frac{k}{n}} {\scr O}_{[n]} \rord (\infty)
	&\mapsfrom
	\oint_{\infty} \frac{d t}{2\pi i}
	t^{k}
	 \left[ \sum_{\ell=0}^\infty A^{(\infty)}_{\ell} t^{-\ell} \right]
	J^a (t) \hat {\scr O}_n(\infty) .
	\label{ModeCovJinf}
\end{align}
On the covering surface the excitations of $\hat{\scr O}_n$ can be read from the OPE
\be	\label{OPEordering}
J^a(t) \hat{\scr O}_n(t_*) = \sum_{k \in {\bb Z}} \frac{\lord J^a_{-k} \hat{\scr O}_n\rord (t_*) }{(t-t_*)^{1-k}} \ ,
\quad
\lord J^a_{-k} \hat{\scr O}_n \rord (t_*) = \oint_{t_*} \frac{d t}{2\pi i} (t-t_*)^{-k} J^a(t) \hat{\scr O}_n(t_*).
\ee
Inserting the OPE in the RHS of Eq.(\ref{ModeCovJ}) we find that the lift of fractional modes on the base produces a \emph{sum} of integer-mode excitations on the covering:
\be	\label{LiftFactroJ}
\lord  J^a_{- \frac{k}{n}} {\scr O}_{[n]}\rord (z_*)  \,
\mapsfrom
\sum_{\ell = 0}^\infty A^{(*)}_{\ell} \lord  J^a_{-k + \ell} \hat {\scr O}_n\rord (t_*) .
\ee
As we will show presently, whenever ${\scr O}_{[n]}$ is NS chiral, anti-chiral or a state in the mutiplet the sum in the RHS is finite.

It is important to note that the ``fractional-mode coefficients'' $A^{(*)}_\ell$ are completely determined from the covering map, and do not depend on the operators apart from their twist structure. More precisely, they depend on the twist structure of the correlation function as whole, which is what defines the covering map.

\subsection{Excitations of NS chiral, anti-chiral and multiplet fields}	\label{SectOnBeingPrimary}

Using formula (\ref{bosonJ3}) for $J^3(t)$ and formulas (\ref{0022NSchiCOVER}) for the lifted NS chirals, 
we can directly write down the OPE
\be	\label{OPEJ3V}
J^3(t) \hat \s^{\frak a}_n(t_*)
	= \frac{q^{\frak a}_n}{t-t_*} \, \hat \s^{\frak a}_n(t_*)
	+ \sum_{\ell = 0}^\infty
	\frac{(t - t_*)^\ell}{\ell!} \lord  (\pa^{\ell} J^3)  \hat \s^{\frak a}_n \rord (t_*) .
\ee
Comparing with Eqs.(\ref{OPEordering}), we see that, for $\ell > 0$,
\be	\label{J0JnVi}
\lord  J^3_\ell \hat \s^{\frak a}_n\rord (t_*) = 0 ,
\quad
\lord  J^3_0 \hat \s^{\frak a}_n\rord (t_*) = q^{\frak a}_n \hat \s^{\frak a}_n(t_*) ,
\quad
\lord  J^3_{-\ell} \hat \s^\fra_n\rord (t_*) 
	= \frac{[{:} (\pa^{\ell-1} J^3) \hat \s^{\frak a}_n {:}](t_*)}{(\ell-1)!} .
\ee
Because the action of all positive modes vanishes, the series in Eq.(\ref{LiftFactroJ}) is finite:
\be	\label{LiftFactroJ2}
\lord  J^3_{- \frac{k}{n}} \s^{\frak a}_{[n]}\rord (z_*)  \,
\mapsfrom
	\sum_{\ell = 0}^k 
	A^{(*)}_{\ell} \lord  J^3_{\ell-k} \hat \s^{\frak a}_n\rord (t_*) ,
\qquad
	k > 0.
\ee

For the $J^\pm$ modes there is also a truncation, but it is different for the different types of fields.
The OPEs
\be	\label{JpmcalV}
\begin{aligned}
	e^{\pm i (\phi_1 - \phi_2)} (t)
	\
	e^{\frac{i}{2} (\rho \phi_1 - \ga \phi_2)}(t_*) 
&
	= 
	(t-t_*)^{\pm\frac{\rho + \ga}2}
	\Bigg[
	e^{\frac{i}{2} [ (\rho \pm 2) \phi_1 - (\ga \pm 2) \phi_2]}(t_*) 
\\
&
	+
	\sum_{\ell =1}^\infty
	\frac{ (t-t_*)^\ell }{\ell!}
	\left(
	{:} \big[\pa^{\ell}  e^{\pm i (\phi_1 - \phi_2)} \big]
	e^{\frac{i}{2} (\rho \phi_1 - \ga \phi_2)}
	{:}
	\right)
	(t_*)
	\Bigg]
\end{aligned}
\ee
summarize the OPE between $J^\pm(t)$ and all NS chirals and anti-chirals. 
For $J^-$, let us focus on the slightly simpler case of $\hat \s^\a_n$ for which $\rho = \ga = n \pm \a$, hence
\be	\label{J0JnViplus}
\begin{aligned}
\lord  J^-_{\ell} \hat \s^\a_n\rord (t_*) 
	&= 0 ,
	&\qquad
	\ell &> n + \a - 1 ,
\\
\lord  J^-_{\ell} \hat \s^\a_n\rord (t_*) 
	&= b_*^{- \nu_{\a}} e^{\frac{i}{2} (n + \a - 2) (\phi_1-\phi_2)}(t_*) ,
	&\qquad
	\ell &= n + \a - 1 ,
\\
\lord  J^-_{\ell} \hat \s^\a_n\rord (t_*) 
	&=
	\frac{
	{:} (\pa^{2(n+\a)-1-\ell}  J^-) \hat \s^\a_n {:}(t_*) 
	}{[2(n + \a) - 1 - \ell]!} ,
	&\qquad
	\ell &< n + \a - 1 .
\end{aligned}
\ee
Note that there is a finite collection of positive modes that do \emph{not} annihilate the fields.
The computation for $\s^{\dot A}_{[n]}$ is analogous, the OPEs can also be read from (\ref{JpmcalV}), with $(J^\pm(t) \hat \s^{\dot A}_n) (t_*) =  (t-t_*)^{\pm n} \big[ O(t-t_*)^0 \big]$.
In particular
\be	\label{VanisOfJminmodes}
\lord  J^-_{\ell} \hat \s^\fra_n\rord (t_*) 
	= 0 ,
	\qquad
	\ell > R_\fra,
\ee
where
\be	\label{DefofRa}
R_{\frak a} =
	\left\{
	\begin{aligned}
	&n +\a -1 &\quad \text{for ${\frak a} = \a$}
	\\
	&n-1&\quad \text{for ${\frak a} = \dot A$}
	\end{aligned}
	\right.
\ee
The NS chiral fields have $ \rho + \ga \geq 2$, hence their OPEs with $J^+$ are regular, as expected. 
More precisely, 
\be	\label{Jpluselliszero}
\lord J^+_\ell \hat \s^\fra_n \rord (t_*) = 0 , \qquad \ell > - (R_\fra +2) .
\ee
Substituting the OPEs into Eq.(\ref{LiftFactroJ}) we find, in accordance with (\ref{Jplusketis0}), that
\be	\label{LiftFactroJAlg}
\lord  J^+_{- \frac{k}{n}} \s^{\frak a}_{[n]}\rord (z_*) 
	= 0,
\qquad 
	\text{for $0 < k < n$} 
\ee
while excitations by $J^-$ produce, as we have claimed, the finite sum 
\be	\label{LiftFactroJ2Moin}
\lord  J^-_{- \frac{k}{n}} \s^{\frak a}_{[n]}\rord (z_*)  \,
	\mapsfrom
	\sum_{\ell = 0}^{k + R_{\frak a}}
	A^{(*)}_{\ell} \lord  J^-_{\ell-k} \hat \s^{\frak a}_n\rord (t_*) .
\ee

For the multiplet fields, assuming $\ell > 0$,
\be	\label{J3Jmrsr}
\begin{aligned}
J^3_\ell \hat \s^{\frak a (r)}_n(t_*)
	=
	{\cal A}^{\fra (r)}_n
	J^3_\ell (J^-_0)^r \hat \s^{\frak a}_n(t_*)
	=
	- r 
	{\cal A}^{\fra (r)}_n
	(J^-_0)^{r-1} J^-_\ell \hat \s^{\frak a}_n(t_*)
\end{aligned}
\ee
This, together with (\ref{VanisOfJminmodes}), allows us to conclude that
\be	\label{VanisOfJ3modesMULTI}
\lord J^3_{\ell} \hat \s^{\fra(r)}_n\rord  
	= 0 
	\qquad \text{for} \qquad
	\ell > R_\fra ,
\ee
and $J^3_{\ell} \hat \s^{\fra(r)}_n  \neq 0$ for $\ell \leq R_\fra$. The non-vanishing operators are, for instance
\be	\label{J0JnViplusMULTI}
\begin{aligned}
\lord  J^3_{\ell} \hat \s^{\a (r)}_n\rord  
	&= 0 ,
	&\qquad
	\ell &> R_\a ,
\\
\lord  J^3_{\ell} \hat \s^{\a(r)}_n\rord  
	&= 
	- \sqrt{ \frac{r}{ 2 q^\a_n - r +1}}
	b_*^{- \nu_{\a}} (J^-_0)^{r-1}  e^{\frac{i}{2} (R_\a - 1) (\phi_1-\phi_2)} ,
	&\qquad
	\ell &= R_\a ,
\\
\lord  J^3_{\ell} \hat \s^{\a(r)}_n\rord  
	&=
	- \sqrt{ \frac{r}{ 2 q^\a_n - r +1}}
	(J^-_0)^{r-1}
	\frac{
	{:} (\pa^{2(n+\a)-1-\ell}  J^-) \hat \s^\a_n {:} 
	}{[2(n + \a) - 1 - \ell]!} ,
	&\qquad
	\ell &< R_\a .
\end{aligned}
\ee
For $\ell = 0$ a computation like (\ref{J3Jmrsr}) gives $( J^3_0 \hat \s^{\frak a(r)}_n ) = (q^{\frak \a}_n - r) \hat \s^{\frak a(r)}_n$, as expected.
Similarly, for $J^-$ modes, 
we have
$(J^-_\ell \hat \s^{\frak a (r)}_n) = {\cal A}^{\fra (r)}_n ( (J^-_0)^{r} J^-_\ell \hat \s^{\frak a}_n)$.
which implies, given (\ref{VanisOfJminmodes}), that 
\be	\label{VanisOfJminmodesMULTI}
\lord  J^-_{\ell} \hat \s^{\fra(r)}_n\rord (t_*) 
	= 0 ,
	\qquad
	\ell > R_\fra,
\ee
and 
$\lord  J^-_{\ell} \hat \s^{\fra(r)}_n\rord  = 0$ for $\ell < R_\fra$, with formulas similar to (\ref{J0JnViplusMULTI}).
Combining everything with Eq.(\ref{J0JnViplus}), we then find, for $k > 0$,
\be	\label{LiftFactroJ2Multi}
\begin{aligned}
\lord  J^3_{- \frac{k}{n}} \s^{\frak a (r)}_{[n]}\rord (z_*)  \,
&\mapsfrom
	- \sqrt{ r (2 q^\fra_n - r )}
	\sum_{\ell = 0}^{k + R_{\frak a}}
	A^{(*)}_{\ell} \lord  J^-_{\ell-k} \hat \s^{\frak a (r-1)}_n\rord (t_*) ,
\\
\lord  J^-_{- \frac{k}{n}} \s^{\frak a (r)}_{[n]}\rord (z_*)  \,
&\mapsfrom
	\sum_{\ell = 0}^{k + R_{\frak a}}
	A^{(*)}_{\ell} \lord  J^-_{\ell-k} \hat \s^{\frak a (r)}_n\rord (t_*) .
\end{aligned}
\ee
A finite sum as claimed.

Finally, let us consider fractional modes of \emph{anti-}chirals fields. 
The OPE (\ref{OPEJ3V}) holds also for anti-chiral fields, but with the opposite sign of $q^{\frak a}_n$, so (\ref{J0JnVi}) also hold:
\be	\label{J0JnViANTI}
\lord  J^3_\ell \hat \s^{\fra \dagger}_n\rord (t_*) = 0 ,
\quad
\lord  J^3_0 \hat \s^{\fra \dagger}_n\rord (t_*) = - q^{\frak a}_n \hat \s^{\fra \dagger}_n(t_*) ,
\quad
\lord  J^3_{-\ell} \hat \s^{\fra \dagger}_n \rord (t_*)
	= \frac{[{:} (\pa^{\ell-1} J^3) \hat \s^{\fra \dagger}_n {:}](t_*)}{(\ell-1)!} ,
\ee
hence
\be	\label{LiftFactroJ2ANTI}
\lord  J^3_{- \frac{k}{n}} \s^{\frak a}_{[n]}\rord (z_*)  \,
\mapsfrom
	\sum_{\ell = 0}^k 
	A^{(*)}_{\ell} \lord  J^3_{\ell-k} \hat \s^{\frak a}_n\rord (t_*) ,
\qquad
	k > 0.
\ee
As for $J^\pm$ modes, for definiteness, let us focus on $\s^{\a\dagger}_{[n]}$. The relevant OPEs are again (\ref{JpmcalV}), now with $\rho = \ga = - (n \pm \a)$, which lead to  
\be	\label{J0JnViplusANTI}
\begin{aligned}
\lord  J^+_{\ell} \hat \s^{\a\dagger}_n\rord (t_*) 
	&= 0 
	& \ell &> R_\a,
\\
\lord  J^+_{\ell} \hat \s^{\a\dagger}_n\rord (t_*) 
	&= b_*^{- \nu_{\a}} e^{-\frac{i}{2} (n + \a - 2) (\phi_1-\phi_2)}(t_*) ,
	& \ell &= R_\a,
\\
\lord  J^+_{\ell} \hat \s^{\a\dagger}_n\rord (t_*) 
	&=
	\frac{
	{:} (\pa^{2(n+\a)-1-\ell}  J^+) \hat \s^{\a\dagger}_n {:}(t_*) 
	}{[2(n + \a) - 1 - \ell]!} ,
	&\qquad
	\ell &< R_\a.
\end{aligned}
\ee
This analogous to (\ref{J0JnViplus}), with $J^+$ replaced by $J^-$. 
This change of $J^+$ and $J^-$ is a common feature. 
We have another finite sum 
\be	\label{LiftFactroJ2MoinANTI}
\lord  J^+_{- \frac{k}{n}} \s^{\fra \dagger}_{[n]}\rord (z_*)  \,
	\mapsfrom
	\sum_{\ell = 0}^{k + R_{\fra}}
	A^{(*)}_{\ell} \lord  J^+_{\ell-k} \hat \s^{\fra \dagger}_n\rord (t_*) .
\ee
Meanwhile, tho OPEs with $J^-$ are regular, that is
\be	\label{JminAntiell0}
\lord J^-_\ell \hat \s^{\fra\dagger}_n \rord (t_*) = 0 , \qquad \ell > - (R_\fra + 2) ,
\ee
and therefore negative-mode excitations of anti-chirals by $J^-$ vanish
\be	\label{LiftFactroJAlgDual}
\lord  J^-_{- \frac{k}{n}} \s^{\frak a \dagger}_{[n]}\rord (z_*) 
	= 0,
\qquad 
	\text{for $k > 0$} 
\ee
which is the anti-chiral version of (\ref{LiftFactroJAlg}).

\bigskip

In computing correlation functions below, we will place excitations of fields at infinity. Recall that, for the point at infinity, the definition of the modes gives
\be	\label{Jaellmodeinfty}
\lord  J^a_\ell \hat {\scr O}\rord (\infty) 
	= 
	\oint_\infty \frac{dt}{2\pi i} t^\ell J^a(t)  \hat {\scr O}(\infty)
	= 
	\oint_{0} \frac{dt'}{2\pi i}t'^{-\ell} J^a(t')  \hat {\scr O}(0)
	= \lord  J^a_{-\ell}  \hat {\scr O}\rord (0)  
\ee
where we made the change of variables $t' = 1/t$. Hence for situations such as (\ref{J0JnVi}), for which all positive-mode excitations of $\hat {\scr O}(0)$ vanish, at infinity it is the \emph{negative} modes who vanish.

\begin{center}
*
\end{center}

In a CFT with current algebra, the usual definition of ``primary'' states are those satisfying (\ref{PrimLJ}). These conditions include those for ``Virasoro primary'', related to the vanishing of positive Virasoro modes, as well as the vanishing of positive current modes.
If we consider only the \emph{integer} mode algebra, the conditions (\ref{PrimLJ}) can be equivalently stated as
\be	\label{PrimLJINTEG}
L_\ell \ket{h,q}_n =0 ,
\quad
J^a_\ell \ket{h,q}_n = 0 
\quad \text{for all $\ell \geq 1 .$}
\ee
The NS chirals, anti-chirals and the multiplet fields, $\s^\fra_{[n]}$, $\s^{\fra \dagger}_{[n]}$, $\s^{\fra(r)}_{[n]}$, all satisfy (\ref{PrimLJINTEG}), so they are primary in this sense.
But if we take into account fractional modes, there is an interesting caveat.

Going back to the OPEs (\ref{J0JnViplus}) we see that, for \emph{positive modes} with $0 < k/n < 1$,
\be	\label{LiftFactroJPRIM}
\lord  J^-_{\frac{k}{n}}\s^{\frak a}_{[n]}\rord (z_*)  \,
\mapsfrom
\sum_{\ell = 0}^\infty A^{(*)}_{\ell} \lord  J^-_{k + \ell} \hat \s^{\frak a}_{n} \rord (t_*) 
=
\sum_{\ell = 0}^{n - 1 + \a - k}  A^{(*)}_{\ell} \lord  J^-_{k + \ell} \hat \s^{\frak a}_{n} \rord (t_*) 
\ee
or, more generally (and rearranging the sum),
\be
\lord  J^-_{\frac{k}{n}} \s^{\frak a}_{[n]}\rord (z_*)  \,
	\mapsfrom
	\sum_{\ell = k}^{R_{\frak a}}
	A^{(*)}_{\ell-k} \lord  J^-_{\ell} \hat \s^{\frak a}_n\rord (t_*) ,
\ee
which do not vanish, that is%
	\footnote{Note that $k$ is an integer, so in fact we are assuming $k \leq n-1$. There is one single exception, namely $\lord J^-_{\frac{n-1}{n}} \s^-_{[n]} \rord = 0$, because the sum of covering surface modes vanishes.}
\be	\label{LiftFactroJAlg}
\lord J^-_{\frac{k}{n}} \s^{\frak a}_{[n]} \rord (z_*) \neq 0,
\qquad 
	\text{for $0 < k < n-1$} .
\ee
Note that for $k = n$, however, we do have $( J^-_1 \s^{\frak a}_{[n]})  = 0$.
Eqs.(\ref{LiftFactroJAlg}) are an interesting type of violation of the set of highest-weight conditions (\ref{PrimLJ}). That is, between the zero mode $J^-_0$ and the would-be-first positive mode $J^-_1$, there is now a collection of $n-1$ positive modes $J^-_{k/n}$ which do not annihilate $\ket{h,h}$. 
The violation (\ref{LiftFactroJAlg}) of the conditions (\ref{PrimLJINTEG}) implies that $J^-_{-k/n}$ excitations of $\s^\fra_{[n]}$ with \emph{negative} modes between zero and one, are \emph{not} Virasoro primary. 
Using the fractional mode algebra, which leads to Eq.(\ref{FracModAlgLJnotprim}), we have
\be
L_1 J^-_{-\frac{k}{n}} \ket{\s^\fra}_n = \frac{k}{n} J^-_{\frac{n - k}{n}} \ket{\s^\fra}_n \neq 0 ,
\qquad
	\text{for}
\quad	 0 < k < n-1 .
\ee

Anti-chiral and multiplet fields suffer from a similar problem. 
From Eqs.(\ref{VanisOfJ3modesMULTI}), (\ref{VanisOfJminmodesMULTI}) and (\ref{Jaellmodeinfty}) it follows that
\be	\label{LiftFactroJPRIMsdm}
\begin{aligned}
\lord  J^-_{\frac{k}{n}} \s^{\frak a (r)}_{[n]}\rord (z_*)  \,
&	\mapsfrom
	\sum_{\ell = k}^{R_{\fra}}
	A^{(*)}_{\ell - k} \lord  J^-_{\ell} \hat \s^{\frak a (r)}_n\rord (t_*) ,
\quad
\lord  J^+_{\frac{k}{n}} \s^{\fra \dagger}_{[n]}\rord (z_*)  \,
	\mapsfrom
	\sum_{\ell = k}^{R_{\fra}}
	A^{(*)}_{\ell - k} \lord  J^+_{\ell} \hat \s^{\fra \dagger}_n\rord (t_*) 
\\
\lord  J^3_{\frac{k}{n}} \s^{\frak a (r)}_{[n]}\rord (z_*)  \,
&	\mapsfrom
	\sum_{\ell = k}^{R_{\fra}}
	- \sqrt{ r (2 q^\fra_n - r )}
	\sum_{\ell = k}^{R_{\frak a}}
	A^{(*)}_{\ell - k} \lord  J^-_{\ell} \hat \s^{\frak a (r-1)}_n\rord (t_*) ,
\end{aligned}
\ee
none of which vanish.
Much like above, the algebra (\ref{ModeAlgebra}) then implies that
\be
\begin{aligned}
L_1 J^-_{-\frac{k}{n}} \ket{\s^{\fra(r)}}_n &= \frac{k}{n} J^-_{\frac{n - k}{n}} \ket{\s^{\fra(r)}}_n \neq 0 ,
\\
L_1 J^3_{-\frac{k}{n}} \ket{\s^{\fra(r)}}_n &= \frac{k}{n} J^3_{\frac{n - k}{n}} \ket{\s^{\fra(r)}}_n \neq 0 ,
\\
L_1 J^+_{-\frac{k}{n}} \ket{\s^{\fra\dagger}}_n &= \frac{k}{n} J^+_{\frac{n - k}{n}} \ket{\s^{\fra\dagger}}_n \neq 0 ,
\qquad
	\text{for}
\quad	 0 < k < n-1 .
\end{aligned}
\ee
Hence \emph{none of these excited states are Virasoro primary.}

Meanwhile, the OPEs (\ref{J0JnVi}) and (\ref{J0JnViANTI}) give us
\be
J^3_{k/n} \ket{\s^\fra}_n = 0,
\qquad 
J^3_{k/n} \ket{\s^{\fra \dagger}}_n = 0,
\qquad
	\text{for}
\quad	 0 < k < n-1 ,
\ee
and the same computation with the mode algebra shows us that the excited fields 
$J^3_{-k/n} \s^\fra_{[n]}$
and
$J^3_{-k/n} \s^{\fra \dagger}_{[n]}$
\emph{are,} indeed, Virasoro primaries. 

\bigskip

Another way of describing the situations is this. 
On the covering surface, we have just one single copy of the $\frak{su}(2)_1$ Kac-Moody algebra (and of the Virasoro algebra). A collection of fields $\Phi^r_{\scr J}(t)$ on $\Scov$ are Kac-Moody primaries if their OPEs with the current is of the form
\be	\label{KacMoodyPrim}
J^a(t) \Phi^r_{\scr J}(t_*) = \frac{ \sum_{r'} [{\scr J}^a]^r{}_{s'} \Phi_{\scr J}^{s'}(t_*) }{t-t_*} + \text{Regular}
\ee
where ${\scr J}^a$ are the matrices of some $\frak{su}(2)$ representation.
This agrees with (\ref{PrimLJINTEG}) for integer modes.
What we have shown above is that NS chiral, anti-chiral and the multiplets do \emph{not} lift to Kac-Moody primaries.
The $J^3$ OPE (\ref{OPEJ3V}) (with a minus sign in the first term for anti-chiral fields) does match (\ref{KacMoodyPrim}), but the OPEs with $J^\pm$ are too singular:
\be	\label{KacMoodyNOT}
\begin{aligned}
 J^-(t) \hat \s^\fra_n  (t_*) &= \frac{V^\fra_n(t_*)}{(t-t_*)^{1+ R_\fra}} + O(t-t_*)^{-R_\fra} ,
\\
 J^+(t) \hat \s^{\fra\dagger}_n  (t_*) &= \frac{V^{\fra\dagger}_n(t_*)}{(t-t_*)^{1+ R_\fra}} + O(t-t_*)^{-R_\fra} ,
\end{aligned}
\ee
where $V^\fra_n$ is some exponential field.
The respective OPEs with $J^+$ and $J^-$ are, instead, too regular, so they do match (\ref{KacMoodyPrim}) with the singularity being absent. 
For multiplets, the OPEs with $J^-$ and with $J^3$ are both too singular.
Thus we see from (\ref{KacMoodyNOT}) that the only NS chiral field that does lift to a primary on the covering is $\s^-_{[2]}$, for which $R_- = 0$, and then $\hat\s^-_2$ and $\hat\s^{-\dagger}_2$ work as a doublet.

\section{Four-point functions}

Eq.(\ref{LiftFactroJ}) implies that a correlation function involving fractional descendants on the base lifts to a \emph{sum} of functions on the covering. 
In the remaining of this paper, we will interested in four-point functions of the form
\be	\label{scrGscrOOO}
G(u)
= 
\Big\langle
\lord  J^{a'}_{\frac{k'}{n'}} {\scr O}'_{[n']}\rord (\infty) \ 
	{\scr O}^1_{[n_1]}(1) \
	{\scr O}^2_{[n_2]}(u) \
	\lord J^a_{-\frac{k}{n}} {\scr O}_{[n]}\rord (0)
\Big\rangle
\ee
where $u \in \Sbase$ is a base-space cross-ratio.
More specifically, we will take the ${\scr O}$ fields in the function to be NS chirals, anti-chirals, multiplet fields, or the interaction operator that deforms the CFT.
In any case, the function lifts to the covering as
${\scr G}(u) \mapsfrom \hat {\scr G}({\bs t})$
where
\be	\label{FungJJoooB}
\hat G({\bs t})=
\sum_{\ell', \ell \geq 0} 
	 A^{(\infty)}_{\ell'}
	 A^{(0)}_{\ell}
	\Big\langle
	\lord  J^{a'}_{k' -\ell'} \hat {\scr O}_{n'}^\infty \rord (\infty) \ 
	\hat {\scr O}^1_{n_1}(t_1) \
	\hat {\scr O}^2_{n_2}(t_2) \
	\lord J^a_{-k + \ell} \hat {\scr O}_{n}\rord (0) 
	\Big\rangle .
\ee
Here we assume that the covering map is such that 
$\{0 \in \Scov\} \mapsto \{ 0 \in \Sbase \}$,
$\{\infty \in \Scov\} \mapsto \{\infty \in \Sbase \}$,
$t_1 \mapsto 1$
and
$t_2 \mapsto u$.
This can always be arranged by making global conformal transformations.%
	\footnote{%
	Branching points in $\Sbase$ may have more than one pre-image in $\Scov$, see \S\ref{Sect4ptBoubleCycle}.}

On the covering surface we have an untwisted CFT, with a single copy of the Virasoro and R-current algebra, and one could hope to express each correlator in the sum (\ref{FungJJoooB}) in terms of a correlator without excitations. 
What we have in mind is something like the fundamental result of BPZ \cite{Belavin:1984vu} that, if $\Phi_i$ are a collection of $Q$ Virasoro primaries with weight $h_i$, then
\be	\label{BPZform}
\begin{aligned}
&
	\big\langle
	\lord L_{-k-1} \Phi_1\rord (t_1)
	\Phi_2(t_2)
	\cdots
	\Phi_Q(t_Q)
	\big\rangle
\\
&\qquad
	=
	\sum_{i=2}^Q
	\left( \frac{k h_i}{(t_i - t_1)^{k+1}} - \frac{1}{(t_i - t_1)^k} \frac{\pa}{\pa t_i} \right)
	\big\langle
	\Phi_1(t_1)
	\Phi_2(t_2)
	\cdots
	\Phi_Q(t_Q)
	\big\rangle .
\end{aligned}
\ee
This is a ``reduction'' of a function with excitations (here Virasoro excitations in the LHS) to a correlator without excitations (the correlator of primaries in the RHS.)
Similar formulas can be derived for more complicated Virasoro excitations, as it is well known \cite{Belavin:1984vu}.
The authors of \cite{Burrington:2022dii} show that this can also be done in the orbifold for Virasoro excitations, when the excited fields lift to Virasoro primaries on the covering.

The counterpart of the functions in the RHS of Eq.(\ref{FungJJoooB}) without the $J^a$ excitations is
\be	\label{FungJJoooBPRIM}
	\Big\langle
	 \hat {\scr O}_{n'}^\infty (\infty) \ 
	\hat {\scr O}^1_{n_1}(t_1) \
	\hat {\scr O}^2_{n_2}(t_2) \
	 \hat {\scr O}_{n}(0)  
	\Big\rangle .
\ee
These have been computed in the literature for many combinations of the fields that are of interest to us, so if we could find a way of reducing (\ref{FungJJoooB}) to a linear operator acting on (\ref{FungJJoooBPRIM}), the former would be, essentially, solved.
For the $\frak{su}(2)_1$ algebra the analogue of relations like (\ref{BPZform}) are the Knizhnik-Zamolodchikov relations \cite{Knizhnik:1984nr}.
The problem is that, as discussed in \S\ref{SectOnBeingPrimary}, neither NS chirals, anti-chirals or multiplet fields lift to Kac-Moody primaries.
It follows that functions in the RHS of (\ref{FungJJoooB}) will not, in general, be amenable to a reduction to (\ref{FungJJoooBPRIM}). 
Functions with $J^3$ excitations involving only chirals and anti-chirals, however, \emph{will} be reducible, because the OPEs of $J^3$ with NS chirals and anti-chirals behave like (\ref{KacMoodyPrim}). 
As for the interaction field, we will show that it \emph{does} lift to a field that behaves as a Kac-Moody primary on the cover. This leads to important simplifications and we will show that there is an important class of functions with $J^\pm$ excitations that can be reduced to (\ref{FungJJoooBPRIM}).

\subsection{Functions with NS chirals and their excitations}	\label{SectFunctNSchiral}

First we take the fields in (\ref{scrGscrOOO}) to be NS chirals, and their relatives: for balancing the total R-charge, the functions must also include NS anti-chiral fields and/or multiplet fields with negative charge. To be specific, we will consider two classes of functions.
The first contains two chiral and two anti-chiral fields, in the form
\be	\label{4ptFunctkknnA}
\Big\langle
\lord  J^{a\dagger}_{\frac{k'}{n'}} \s^{\frak a' \dagger}_{[n']}\rord (\infty) 
\; \s^{\frak b_1 \dagger}_{[n_1]}(1)
\; \s^{\frak b_2}_{[n_2]}(u)
\lord J^a_{-\frac{k}{n}} \s^{\frak a}_{[n]} \rord (0)
\Big\rangle .
\ee
The second contains one multiplet field:
\be	\label{4ptFunctkknnB}
\Big\langle
\lord  J^{a\dagger}_{\frac{k'}{n'}} \s^{\frak a' \dagger}_{[n']}\rord (\infty) 
\; \s^{\frak b_1 (r)}_{[n_1]}(1)
\; \s^{\frak b_2}_{[n_2]}(u)
\lord J^a_{-\frac{k}{n}} \s^{\frak a}_{[n]} \rord (0)
\Big\rangle .
\ee
In both cases, we assume $k, k' \geq 1$, and all twists are non-trivial.
It will often be convenient to use the following abbreviated notation  as a shorthand for both (\ref{4ptFunctkknnA}) and (\ref{4ptFunctkknnB}):
\be	\label{4ptFunctkknn}
G_\NS(u)
= 
\Big\langle
\lord  J^{a\dagger}_{\frac{k'}{n'}} \s^{\frak a' \dagger}_{[n']}\rord (\infty) 
\; \s^{\frak b_1 \sim}_{[n_1]}(1)
\; \s^{\frak b_2}_{[n_2]}(u)
\lord J^a_{-\frac{k}{n}} \s^{\frak a}_{[n]} \rord (0)
\Big\rangle ,
\ee
where $\sim$ in $\hat \s^{\frb_2 \sim}_{n_2}$ indicates a dagger or a multiplet.

According to Eq.(\ref{FungJJoooB}), $G_\NS({\bs z})$ lifts to the covering surface as a sum of correlators with integer excitations:
\bsub	\label{Gtsumell}
\begin{flalign}
&&\hat G^{\{a; \fra_i ; \frb_i\}}_\NS({\bs t}) 
	= 
	\sum_{\ell = - R_{\fra}}^{k} 
	\sum_{\ell' = -R_\fra'}^{k'}
	A^{(\infty)}_{k'-\ell'}
	A^{(0)}_{k-\ell}
	\hat G^\NS_{\ell',\ell}({\bs t})
&&
\\
\text{where}
&&
\hat G^\NS_{\ell',\ell}({\bs t})
	=
	\Big\langle
	\lord  J^{a\dagger}_{\ell'} \hat \s^{\fra' \dagger}_{n'} \rord (\infty) 
	\; \hat \s^{\frb_1 \sim}_{n_1}(t_1) 
	\; \hat \s^{\frb_2}_{n_2}(t_2) \;
	\lord J^a_{- \ell} \hat \s^\fra_n \rord (0)
	\Big\rangle
&&
	\label{GtsumellB}
\end{flalign}
\esub
Here $R_\fra$, $R_\fra'$ are given by (\ref{DefofRa}) for $J^\pm$ excitations, and we define $R_\fra = R_\fra' = 0$ for $J^3$ excitations, to be consistent with (\ref{LiftFactroJ2}).

The idea behind BPZ and Knizhnik-Zamolodchikov relations is use Ward identities to manipulate the excitation modes through the fields in the correlator. The identities we will use are derived in App.\ref{AppWardId}.
We now move the non-negative modes from the field at infinity to the other fields in the correlator using the Ward identities (\ref{JpWardId}).
For $\ell' \geq 0$, these give
\be	\label{GNSalge}
\begin{aligned}
\hat G^\NS_{\ell',\ell}({\bs t})
&	=
	-
	\Big\langle
	  \hat \s^{\fra' \dagger}_{n'}  (\infty) 
	\; \hat \s^{\frb_1 \sim}_{n_1}(t_1) 
	\; \hat \s^{\frb_2}_{n_2}(t_2) \;
	 [J^{a\dagger}_{\ell'} , J^{a}_{- \ell} ] \hat \s^\fra_n  (0)
	\Big\rangle
\\
&\quad \,
	-
	\Big\langle
	\hat \s^{\fra' \dagger}_{n'}  (\infty) 
	\; \hat \s^{\frb_1 \sim}_{n_1}(t_1) 
	\; \hat \s^{\frb_2}_{n_2}(t_2) \;
	\lord  J^{a}_{- \ell} \lord J^{a\dagger}_{\ell'}  \hat \s^\fra_n \rord\rord  (0)
	\Big\rangle
\\
&\quad \,
	-
	\sum_{s' = 0}^{\ell'}
	{\ell' \choose s'} \,
	t_1^{\ell' - s'}
	\Big\langle
	\lord  \hat \s^{\fra '\dagger}_{n'} \rord (\infty) 
	\; \lord J^{a\dagger}_{s'} \hat \s^{\frb_1 \sim}_{n_1} \rord (t_1) 
	\; \hat \s^{\frb_2}_{n_2}(t_2) \;
	\lord  J^a_{- \ell} \hat \s^\fra_n \rord (0)
	\Big\rangle
\\
&\quad \,
	-
	\sum_{s' = 0}^{\ell'}
	{\ell' \choose s'} \,
	t_2^{\ell' - s'}
	\Big\langle
	\hat \s^{\fra' \dagger}_{n'}  (\infty) 
	\;  \hat \s^{\frb_1 \sim}_{n_1}(t_1) 
	\; \lord J^{a\dagger}_{s'} \hat \s^{\frb_2}_{n_2} \rord(t_2) \;
	\lord  J^a_{- \ell} \hat \s^\fra_n \rord (0)
	\Big\rangle ,
\end{aligned}
\ee
Here we can see the rationale behind using the Ward identities. Most of the terms in the RHS of the previous equations have \emph{positive} modes acting on fields. \emph{If} these (covering-surface) fields were Kac-Moody primary, these terms would vanish. 
But this does not happen in general. Now it is more convenient to consider different cases separately.

\bigskip

\noindent
{\bf $J^3$ modes.}
\\
For $J^a = J^{a\dagger} = J^3$ all functions have $\ell,\ell' \geq 0$ because $R_\fra = R_\fra' = 0$.
Eq.(\ref{GNSalge}) becomes
\be	\label{GNSalgeJ3}
\begin{aligned}
\hat G^\NS_{\ell',\ell}({\bs t})
&	=
	-
	\frac{\ell}{2} \delta_{\ell',\ell}
	\Big\langle
	\hat \s^{\fra' \dagger}_{n'}  (\infty) 
	\; \hat \s^{\frb_1 \sim}_{n_1}(t_1) 
	\; \hat \s^{\frb_2}_{n_2}(t_2) \;
	\hat \s^\fra_n (0) 	\Big\rangle
\\
&\quad \,
	-
	\delta_{\ell',0} \, q^\fra_n \
	\Big\langle
	\hat \s^{\fra' \dagger}_{n'}  (\infty) 
	\; \hat \s^{\frb_1 \sim}_{n_1}(t_1) 
	\; \hat \s^{\frb_2}_{n_2}(t_2) \;
	\lord  J^{3}_{- \ell} \hat \s^\fra_n \rord (0)
	\Big\rangle
\\
&\quad \,
	-
	\sum_{s' = 0}^{\ell'}
	{\ell' \choose s'} \,
	t_1^{\ell' - s'}
	\Big\langle
	\hat \s^{\fra' \dagger}_{n'}  (\infty) 
	\; \lord J^{3}_{s'} \hat \s^{\frb_1 \sim}_{n_1}\rord (t_1) 
	\; \hat \s^{\frb_2}_{n_2}(t_2) \;
	\lord  J^3_{- \ell} \hat \s^\fra_n \rord (0)
	\Big\rangle
\\
&\quad \,
	-
	q^{\frb_2}_{n_2} t_2^{\ell'}
	\Big\langle
	\hat \s^{\fra' \dagger}_{n'}  (\infty) 
	\;  \hat \s^{\frb_1 \sim}_{n_1}(t_1) 
	\; \hat \s^{\frb_2}_{n_2}(t_2) \;
	\lord  J^3_{- \ell} \hat \s^\fra_n \rord (0)
	\Big\rangle
\end{aligned}
\ee
We have used that $[J^3_{\ell'}, J^3_{-\ell} ] = \frac12 \ell \delta_{\ell',\ell}$, and that $J^3_\ell \hat \s^\frc_n = 0$ for $\ell > 0$.

Take $\hat\s^{\frb_1 \sim}_{n_1} = \hat\s^{\frb_1 \dagger}_{n_1}$ to be an anti-chiral field.
Since $J^3_{s'} \hat \s^{\frb_1 \dagger}_{n_1} = 0$ for all $s' > 0$, only the zero mode contributes (with a multiplicative factor) in the sum in the RHS of (\ref{GNSalgeJ3}).
Now we use the Ward identity (\ref{JpWardIdfin}) to exchange the mode $J^3_{-\ell}$ at zero by modes acting on the other fields. The action of the modes on all fields vanishes except for $\ell = 0$, and we get
\be	\label{GNSalgeJ3BANTI}
\begin{aligned}
\hat G^\NS_{\ell',\ell}({\bs t})
&	=
	B_{\ell', \ell}({\bs t}) \
	\Big\langle
	\hat \s^{\fra' \dagger}_{n'}  (\infty) 
	\; \hat \s^{\frb_1 \dagger}_{n_1}(t_1) 
	\; \hat \s^{\frb_2}_{n_2}(t_2) \;
	\hat \s^\fra_n (0) 	\Big\rangle
\\
B_{\ell', \ell}({\bs t})
	&=
	- \Big[
	\frac{\ell}{2} \; \delta_{\ell,\ell'}
	+
	\big( q^{\fra}_{n} \delta_{\ell',0} - q^{\frb_1}_{n_1} t_1^{\ell'} + q^{\frb_2}_{n_2} t_2^{\ell'} \big)
	\big( q^{\fra'}_{n'} \delta_{\ell,0} + q^{\frb_1}_{n_1} t_1^{-\ell} - q^{\frb_2}_{n_2} t_2^{-\ell} \big)
	\Big]
\end{aligned}
\ee
The entire dependence on the mode excitations is on the functions $B_{\ell',\ell}$, so Eq.(\ref{Gtsumell}) becomes
\be	\label{GNS3scrD}
\hat G^{\{3;\fra_i;\frb_i \}}_\NS({\bs t}) 
	= 
	{\scr D}^{\{3;k;\fra_i;\frb_i \}}_\NS({\bs t}) \
	\Big\langle
	\hat \s^{\fra' \dagger}_{n'}  (\infty) 
	\; \hat \s^{\frb_1 \dagger}_{n_1}(t_1) 
	\; \hat \s^{\frb_2}_{n_2}(t_2) \;
	\hat \s^\fra_n (0) 	\Big\rangle
\ee
where 
\be	\label{ScrD3antNS}
{\scr D}^{\{3;\fra_i;\frb_i \}}_\NS({\bs t})
	=
	\sum_{\ell = 0}^{k} 
	\sum_{\ell' = 0}^{k'}
	A^{(\infty)}_{k'-\ell'}
	A^{(0)}_{k-\ell}
	B_{\ell', \ell}({\bs t})
\ee
We have reduced the function on the covering surface to a function without excitations. That is, the function (\ref{ScrD3antNS}) is completely computable from the $B$-coefficients in (\ref{GNSalgeJ3BANTI}) and the $A$-coefficients (\ref{AstaellDerccBell}) (assuming the covering map is given).
As mentioned above, the correlators without excitations have been computed in the literature (see e.g.~\cite{Lima:2022cnq}), so the function with excitations is now given as well, by Eq.(\ref{GNS3scrD}).

\bigskip

On the other hand, correlation functions containing multiplet fields \emph{cannot} be reduced to functions containing only unexcited fields.
For multiplet fields, the sum over $s$ in (\ref{GNSalgeJ3}) has more terms. Moving the mode $J^a_{-\ell}$ from $\hat \s^\fra_n$ we arrive at
{\small
\be	\label{GNSalgeJ3BMULTI}
\begin{aligned}
&\hat G^\NS_{\ell',\ell}({\bs t})
	=
\\
&
	\left(
	- \frac{\ell}{2} \delta_{\ell',\ell}
	+
	\left[ \delta_{\ell',0} \, q^\fra_n + q^{\frb_2}_{n_2} t_2^{\ell'} \right]
	\left[ q^{\frb_2}_{n_2} - q^{\fra'}_{n'} \right] \delta_{\ell,0}
	\right)
	\Big\langle
	\hat \s^{\fra' \dagger}_{n'}  (\infty) 
	\;  \hat \s^{\frb_1 (r)}_{n_1}(t_1) 
	\; \hat \s^{\frb_2}_{n_2}(t_2) \;
	\hat \s^\fra_n (0)
	\Big\rangle
\\
&
	+
	\sum_{s = 0}^{R_{\frb_1}} (-)^s {\ell + s-1 \choose s} t_1^{s-\ell}
	\left[
	\delta_{\ell',0} \, q^\fra_n 
	+ q^{\frb_2}_{n_2} t_2^{\ell'}
	\right]
	\Big\langle
	\hat \s^{\fra' \dagger}_{n'}  (\infty) 
	\; \lord J^{3}_{s}  \hat \s^{\frb_1 (r)}_{n_1}\rord (t_1) 
	\; \hat \s^{\frb_2}_{n_2}(t_2) \;
	\hat \s^\fra_n (0) 	\Big\rangle
\\
&
	+
	\sum_{s = 0}^{R_{\frb_1}}
	\sum_{s' = 0}^{\ell'}
	(-)^s {\ell + s-1 \choose s}
	{\ell' \choose s'} \,
	t_1^{s-s' - \ell + \ell'}
	\Big\langle
	\hat \s^{\fra' \dagger}_{n'}  (\infty) 
	\; \lord J^3_{s'} \lord J^{3}_{s} \hat \s^{\frb_1 (r)}_{n_1}\rord\rord(t_1) 
	\; \hat \s^{\frb_2}_{n_2}(t_2) \;
	\hat \s^\fra_n (0) 	\Big\rangle
\end{aligned}
\ee
}%
The correlators on the second and the last lines cannot be reduced to a correlator of primaries only. We can try to take the excitations out of $J^{3}_{s} \hat \s^{\frb_1 (r)}_{n_1}$ and $J^3_{s'} J^{3}_{s} \hat \s^{\frb_1 (r)}_{n_1}$ by using the Ward identities again, but this does not work because it produces an excitation of the field at infinity. 
For example, the correlator in the second line of (\ref{GNSalgeJ3BMULTI}) can be expressed as
\be
\begin{aligned}
&	\Big\langle
	\hat \s^{\fra' \dagger}_{n'}  (\infty) 
	\;  \lord J^{3}_{s}  \hat \s^{\frb_1 (r)}_{n_1} \rord (t_1) 
	\; \hat \s^{\frb_2}_{n_2}(t_2) \;
	\hat \s^\fra_n (0) 	\Big\rangle
\\
&\qquad\qquad
	=
	- \Big[
	q^{\frb_2}_{n_2}
	(t_1 - t_2)^{s}
	+
	q^{\fra}_{n}
	t_1^{s}
	\Big]
	\Big\langle
	\hat \s^{\fra' \dagger}_{n'}  (\infty) 
	\; \hat \s^{\frb_1 (r)}_{n_1}(t_1) 
	\; \hat \s^{\frb_2}_{n_2}(t_2) \;
	\hat \s^\fra_n (0) 	\Big\rangle
\\
&\qquad\qquad\quad
- 	\Big\langle
	\lord  J^3_s \hat \s^{\fra' \dagger}_{n'} \rord (\infty) 
	\; \hat \s^{\frb_1 (r)}_{n_1}(t_1) 
	\; \hat \s^{\frb_2}_{n_2}(t_2) \;
	  \hat \s^\fra_n (0)
	\Big\rangle .
\end{aligned}
\ee
An excitation remains in the last line, $( J^3_s \hat \s^{\fra' \dagger}_{n'} ) (\infty)  = ( J^3_{-s} \hat \s^{\fra' \dagger}_{n'} )(0) \neq 0$ for $s > 0$.

\bigskip

\noindent
{\bf$J^\pm$ modes.}
\\
Functions with fractional modes of $J^\pm$ are more complicated.
Now $R_\fra, R_\fra' > 0$ so the sum (\ref{Gtsumell}) includes both positive and negative modes.
Let us consider first $\ell' > 0$, whereas Eq.(\ref{GNSalge}) becomes
\be	\label{GNSalgeMIN}
\begin{aligned}
\hat G^\NS_{\ell',\ell}({\bs t})
&	=
	\Big\langle
	\lord  J^{+}_{\ell'} \hat \s^{\fra' \dagger}_{n'} \rord (\infty) 
	\; \hat \s^{\frb_1 \sim}_{n_1}(t_1) 
	\; \hat \s^{\frb_2}_{n_2}(t_2) \;
	\lord J^-_{- \ell} \hat \s^\fra_n \rord (0)
	\Big\rangle
\\
&	=
	-
	\ell' \, \delta_{\ell, \ell'}
	\Big\langle
	\hat \s^{\fra' \dagger}_{n'}  (\infty) 
	\; \hat \s^{\frb_1 \sim}_{n_1}(t_1) 
	\; \hat \s^{\frb_2}_{n_2}(t_2) \;
	\hat \s^\fra_n (0) 	\Big\rangle
\\
&\quad\,
	- 2
	\Big\langle
	\hat \s^{\fra' \dagger}_{n'}  (\infty) 
	\; \hat \s^{\frb_1 \sim}_{n_1}(t_1) 
	\; \hat \s^{\frb_2}_{n_2}(t_2) \;
	\lord J^3_{\ell'-\ell} \hat \s^\fra_n \rord (0)
	\Big\rangle
\\
&\quad \,
	-
	\sum_{s' = 0}^{\ell'}
	{\ell' \choose s'} \,
	t_1^{\ell' - s'}
	\Big\langle
	\hat \s^{\fra' \dagger}_{n'}  (\infty) 
	\; J^{+}_{s'} \hat \s^{\frb_1 \sim}_{n_1}(t_1) 
	\; \hat \s^{\frb_2}_{n_2}(t_2) \;
	\lord  J^-_{- \ell} \hat \s^\fra_n \rord (0)
	\Big\rangle
\end{aligned}
\ee
We can try to simplify this further,
\be	\label{GNSalgeMIN}
\begin{aligned}
&\hat G^\NS_{\ell',\ell}({\bs t}) =
\\
&
	-
	\ell' \, \delta_{\ell, \ell'}
	\Big\langle
	\hat \s^{\fra' \dagger}_{n'}  (\infty) 
	\; \hat \s^{\frb_1 \sim}_{n_1}(t_1) 
	\; \hat \s^{\frb_2}_{n_2}(t_2) \;
	\hat \s^\fra_n (0) 	\Big\rangle
\\
&
	- 2 \theta(\ell- \ell')
	\sum_{s \geq 0}
	(-)^s
	{\ell - \ell' + s - 1 \choose s}
	t_1^{\ell - \ell' -s}
	\Big\langle
	\hat \s^{\fra' \dagger}_{n'}  (\infty) 
	\; \lord J^3_{s} \hat \s^{\frb_1 \sim}_{n_1}\rord (t_1) 
	\; \hat \s^{\frb_2}_{n_2}(t_2) \;
	\hat \s^\fra_n (0) 	\Big\rangle
\\
&
	-
	\sum_{s' = 0}^{\ell'}
	{\ell' \choose s'} \,
	t_1^{\ell' - s'}
	\Big\langle
	\hat \s^{\fra' \dagger}_{n'}  (\infty) 
	\; \lord J^{+}_{s'} \hat \s^{\frb_1 \sim}_{n_1} \rord (t_1) 
	\; \hat \s^{\frb_2}_{n_2}(t_2) \;
	\lord  J^-_{- \ell} \hat \s^\fra_n \rord (0)
	\Big\rangle
\end{aligned}
\ee
It is not difficult to convince oneself that now if $\hat \s^{\frb_2 \sim}_{n_2}$ is \emph{either} anti-chiral \emph{or} a multiplet, the functions \emph{cannot} be fully reduced to correlators involving primaries only. 
If $\hat \s^{\frb_2 \sim}_{n_2} = \hat \s^{\frb_2 \dagger}_{n_2}$ is anti-chiral, the correlators in the second line vanish (except for $s = 0$), but the ones in the last line do not. 
If $\hat \s^{\frb_2 \sim}_{n_2} = \hat \s^{\frb_2 (r)}_{n_2}$ is a multiplet, neither of the correlators vanish.
These non-vanishing correlators can be expressed in different ways with Ward identities, but some excited field always remains.

\subsection{Functions with the deformation operator}

We will now turn to a different class of functions, with two insertions of the deformation operator that drives the D1-D5 CFT away from the free orbifold point: 
\be	\label{Gakint}
G^{\{a;k\}}_{\rm{int}}(u) = 
	\Big\langle
	\lord  J^{a \dagger}_{\frac{k}{n}} \s^{\fra \dagger}_{[n]}\rord (\infty) 
	\ \Oint(1) 
	\ \Oint(u)
	\ \lord J^a_{-\frac{k}{n}} \s^\fra_{[n]} \rord (0)
	\Big\rangle
\ee
Note that now we take conjugate fields at zero and infinity, with the same twist and fractional mode.
These functions can be used for a perturbative computation of the lifting of the excited states $J^a_{-k/n} \ket{\s^\fra}_n$ at second order in conformal perturbation theory, and they also contain structural information about the fusion rules of $\Oint$ with the excited fields.

The interaction operator is one of the SUGRA moduli (see e.g.~\cite{Avery:2010er,Avery:2010qw}), an $S_N$-invariant singlet of the SU(2) symmetries, and it is marginal, with dimension $(h , \tilde h) = (1,1)$. 
It is obtained as a supercurrent excitation of $\s^{+\dot+}_{[2]}$,   
\be \label{DeformwithMo}
\begin{aligned}
\Oint(z_* , \bar z_*) 
	&= \e_{A  B} G^{- A}_{-\frac{1}{2}} \tilde G^{ \dot-  B}_{-\frac{1}{2}} 
	\s^{+\dot+}_{[2]}(z_*, \bar z_*) 	
\\
	&=  \e_{A  B} 
	\oint \frac{d z}{2\pi i}
	\oint \frac{d\bar z}{2\pi i}
	G^{- A}(z) \tilde G^{ \dot-  B}(\bar z) 
	\s^{+\dot+}_{[2]}(z_*, \bar z_*) 	
\end{aligned}
\ee
On the covering surface, it lifts to 
\be
\Oincov(t_*,\bar t_*) \mapsto \Oint(z_*, \bar z_*)
\ee
which can be conveniently expressed as
\be	\label{InteraOpera}
\begin{aligned}
& \Oincov(t_*,\bar t_*)  
\\
&\quad	=
		|b_*|^{- \frac54} \Big[
		 {:} \pa X^{\dot 1 1} \, e^{+\frac{i}{2} (\phi_1 + \phi_2)} 
			\left( \bar \pa X^{\dot 1 2} e^{+\frac{i}{2} (\tilde \phi_1 + \tilde \phi_2)} 
				- (\bar \pa X^{\dot 1 1})^\dagger e^{- \frac{i}{2} (\tilde \phi_1 + \tilde \phi_2)} 
			\right) {:}
\\
&\quad\quad	
			- {:} \pa X^{\dot 1 2} \, e^{+\frac{i}{2} (\phi_1 + \phi_2)} 
			\left( (\bar \pa X^{\dot 1 2})^\dagger e^{- \frac{i}{2} (\tilde \phi_1 + \tilde \phi_2)} 
				+ \bar \pa X^{\dot 1 1} e^{+ \frac{i}{2} (\tilde \phi_1 + \tilde \phi_2)} 
			\right) {:}
\\
&\quad\quad
		+ 	{:} (\pa X^{\dot 1 1})^\dagger e^{- \frac{i}{2} (\phi_1 + \phi_2)} 
			\left( (\bar \pa X^{\dot 1 2})^\dagger e^{-\frac{i}{2} (\tilde \phi_1 + \tilde \phi_2)} 
				+ \bar \pa X^{\dot 1 1} e^{+\frac{i}{2} (\tilde \phi_1 + \tilde \phi_2)} 
			\right) {:}
\\
&\quad\quad
		+ 	{:} (\pa X^{\dot 1 2})^\dagger \, e^{-\frac{i}{2} (\phi_1 + \phi_2)} 
			\left( \bar \pa X^{\dot 1 2} e^{+ \frac{i}{2} (\tilde \phi_1 + \tilde \phi_2)} 
				- (\bar \pa X^{\dot 1 1})^\dagger e^{- \frac{i}{2} (\tilde \phi_1 + \tilde \phi_2)} 
			\right) {:}
			\Big] .
\end{aligned}
\ee  
This formula is what results from computing the contour integrals from (\ref{DeformwithMo}) in the covering surface, see \cite{Burrington:2012yq}. 
Since the bosons $\pa X^{\dot A B}$ are independent from the bosonized fermions $\phi_i$, they behave like spectators of the action of $J^a$ modes on $\Oincov$, which is easily computed from special cases of the OPEs with exponentials that we have seen above, such as (\ref{JpmcalV}) with $\rho = - \ga = \pm 1$. All in all, 
\be	\label{J3ellOincov0}
\lord  J^3_\ell  \Oincov\rord  = 0 , 
\quad
\lord  J^\pm_\ell \Oincov \rord  = 0 
\qquad \text{for $\ell \geq 0$} ,
\ee
but $J^3_\ell \Oincov \neq 0$ and  $( J^\pm_\ell \Oincov ) \neq 0$ for $\ell < 0$. 
These equations mean that $\Oincov$ \emph{does} transform as in Eq.(\ref{KacMoodyPrim}), i.e.~it is a Kac-Moody primary in the trivial (scalar) irrep of $\frak{su}(2)$.
As we will now see, this leads to interesting simplifications in correlation functions.

The function (\ref{Gakint}) lifts to
\bsub	\label{GtsumellINT}
\begin{flalign}
&&\hat G^{\{a;k\}}_\rm{int}({\bs t}) 
	= 
	\sum_{\ell = - R_{\fra}}^{k} 
	\sum_{\ell' = -R_\fra}^{k}
	A^{(\infty)}_{k-\ell'}
	A^{(0)}_{k-\ell}
	\hat G^{\{a;k\}}_{\ell',\ell}({\bs t})
&&
\\
\text{where}
&&
\hat G^{\{a;k\}}_{\ell',\ell}({\bs t})
	=
	\Big\langle
	\lord  J^{a\dagger}_{\ell'} \hat \s^{\fra \dagger}_{n} \rord (\infty) 
	\Oincov(t_1) 
	\Oincov(t_2) \;
	\lord J^a_{- \ell} \hat \s^\fra_n \rord (0)
	\Big\rangle .
&&
	\label{GtsumellB}
\end{flalign}
\esub
For $J^3$ modes, $R_\fra = 0$ and we have $\ell,\ell' \geq 0$.
We can move the modes using Ward identities, in a computation much like (\ref{GNSalgeJ3}). Given (\ref{J3ellOincov0}), the result is the same as (\ref{GNSalgeJ3BANTI}) if we set $q^{\frb_i}_{n_i} = 0$,
\be	\label{GIntalgeJ3}
\begin{aligned}
&\hat G^{\{3;k\}}_{\ell',\ell}({\bs t})
	=
	B_{\ell',\ell}({\bs t}) \
	\Big\langle
	\hat \s^{\fra \dagger}_{n} (\infty) 
	\;  \Oincov(t_1) 
	\; \Oincov(t_2) \;
	\hat \s^\fra_n (0) 	\Big\rangle
\end{aligned}
\ee
where
\be
B_{\ell',\ell}({\bs t}) 
	=
	- \frac{\ell}{2} \delta_{\ell',\ell}
	-
	\delta_{\ell,0} \delta_{\ell',0} \, (q^\fra_n)^2 
\ee
It follows that we have, again, a reduction to a function without excitations:
\be
\hat G^{\{3;k\}}_{\rm{int}}({\bs t}) 
	= {\scr D}^{\{3;k\}}_\rm{int} ({\bs t}) \, 
	\Big\langle
	\hat \s^{\fra \dagger}_{n} (\infty) 
	\;  \Oincov(t_1) 
	\; \Oincov(t_2) \;
	\hat \s^\fra_n (0) 	\Big\rangle
\ee
where, as in (\ref{ScrD3antNS}), 
\be	\label{J3sigmaINT}
\begin{aligned}
{\scr D}^{\{3;k\}}_{\rm{int}}({\bs t})
&	=
	\sum_{\ell = 0}^{k} 
	\sum_{\ell' = 0}^{k'}
	A^{(\infty)}_{k'-\ell'}
	A^{(0)}_{k-\ell}
	B_{\ell', \ell}({\bs t})
	= 
	- \left[
	(q^\fra_n)^2
	A_{k}^{(0)} A_{k}^{(\infty)}
	+
	\sum_{\ell=1}^k 
	\frac{\ell}{2} A_{k-\ell}^{(0)} A_{k-\ell}^{(\infty)}
	\right] 
\end{aligned}
\ee

For $J^\pm$ modes, we can organize the sum (\ref{GtsumellINT}) as 
\be	\label{GtsumellINTmin}
\begin{aligned}
\hat G^{\{-;k\}}_\rm{int}({\bs t}) 
&	= 
	\sum_{\ell = 1}^{R_\fra}
	\sum_{\ell' = 0}^{k} 
	A^{(\infty)}_{k-\ell'}
	A^{(0)}_{k+\ell}
	\hat G^{\{-;k\}}_{\ell',-\ell}({\bs t})
	+
	\sum_{\ell = 0}^{k} 
	\sum_{\ell' = 1}^{R_\fra}
	A^{(\infty)}_{k+\ell'}
	A^{(0)}_{k-\ell}
	\hat G^{\{-;k\}}_{-\ell',\ell}({\bs t})
\\
&
	+
	\sum_{\ell, \ell' = 1}^{R_\fra}
	A^{(\infty)}_{k+\ell'}
	A^{(0)}_{k+\ell}
	\hat G^{\{-;k\}}_{-\ell',-\ell}({\bs t}) 
	+
	\sum_{\ell, \ell' = 0}^{k} 
	A^{(\infty)}_{k-\ell'}
	A^{(0)}_{k-\ell}
	\hat G^{\{-;k\}}_{\ell',\ell}({\bs t}) .
\end{aligned}
\ee
Let us consider the correlation functions one by one. 
First, the function with one negative index
$\hat G^{\{-;k\}}_{\ell',-\ell}$.
Using Ward identities (\ref{JpWardId}) together with (\ref{J3ellOincov0}),
\be	\label{GIntalgeJminA}
\begin{aligned}
\hat G^{\{-;k\}}_{\ell',-\ell}({\bs t})
&	=
	\Big\langle
	\lord  J^{+}_{\ell'} \hat \s^{\fra \dagger}_{n} \rord (\infty) 
	\Oincov(t_1) 
	\Oincov(t_2) \;
	\lord J^-_{\ell} \hat \s^\fra_n \rord (0)
	\Big\rangle 
\\
&	=
	-
	\Big\langle
	\hat \s^{\fra \dagger}_{n} (\infty) 
	\Oincov(t_1) 
	\Oincov(t_2) \;
	\lord  J^{+}_{\ell'}  \lord J^-_{\ell} \hat \s^\fra_n \rord \rord (0)
	\Big\rangle 
\\
&	=
	- 2
	\Big\langle
	\hat \s^{\fra \dagger}_{n} (\infty) 
	\Oincov(t_1) 
	\Oincov(t_2) \;
	\lord  J^{3}_{\ell + \ell'} \hat \s^\fra_n \rord (0)
	\Big\rangle 
\\
	&= 0 
\end{aligned}
\ee
where we used the algebra in the line next-to-last and (\ref{J0JnVi}) in the end. 
Similarly,
\be	\label{GIntalgeJminB}
\begin{aligned}
\hat G^{\{-;k\}}_{-\ell',\ell}({\bs t})
&	=
	\Big\langle
	\lord  J^{+}_{-\ell'} \hat \s^{\fra \dagger}_{n} \rord (\infty) 
	\Oincov(t_1) 
	\Oincov(t_2) \;
	\lord J^-_{-\ell} \hat \s^\fra_n \rord (0)
	\Big\rangle 
\\
&	=
	- 2
	\Big\langle
	\hat \s^{\fra \dagger}_{n} (\infty) 
	\Oincov(t_1) 
	\Oincov(t_2) \;
	\lord  J^{3}_{-\ell - \ell'} \hat \s^\fra_n \rord (0)
	\Big\rangle 
\\
&	=
	2
	\Big\langle
	\lord  J^{3}_{-\ell - \ell'} \hat \s^{\fra \dagger}_{n} \rord (\infty) 
	\Oincov(t_1) 
	\Oincov(t_2) \;
	\hat \s^\fra_n (0) 	\Big\rangle 
\\
	&= 0 
\end{aligned}
\ee
because 
$J^{3}_{-\ell - \ell'} \hat \s^{\fra \dagger}_{n} (\infty) 
	= J^{3}_{\ell + \ell'} \hat \s^{\fra \dagger}_{n} (0)
	= 0 .
$
Note that here we used the Ward identities twice.
Also, to get to the second line, we must recall that $J^+_{-\ell} \hat \s^{\fra \dagger}_{n}(0) = 0$ for $\ell < R_\fra + 2$, as per Eq.(\ref{Jpluselliszero}), a condition that is satisfied in all terms of the sum (\ref{GtsumellINTmin}).
Going to the second line of (\ref{GtsumellINTmin}), we have
\be	\label{GIntalgeJminC}
\begin{aligned}
\hat G^{\{-;k\}}_{-\ell',-\ell}({\bs t})
&	=
	\Big\langle
	\lord  J^{+}_{-\ell'} \hat \s^{\fra \dagger}_{n} \rord (\infty) 
	\Oincov(t_1) 
	\Oincov(t_2) \;
	\lord J^-_{\ell} \hat \s^\fra_n \rord (0)
	\Big\rangle 
\\
&	=
	- 2
	\Big\langle
	\hat \s^{\fra \dagger}_{n} (\infty) 
	\Oincov(t_1) 
	\Oincov(t_2) \;
	\lord  J^{3}_{\ell - \ell'} \hat \s^\fra_n \rord (0)
	\Big\rangle 
\\
&	=
	- 2 q^\fra_n \delta_{\ell,\ell'}
	\Big\langle
	\hat \s^{\fra \dagger}_{n} (\infty) 
	\Oincov(t_1) 
	\Oincov(t_2) \;
	\hat \s^\fra_n (0) 	\Big\rangle 
\\
&	=
	\hat G^{\{-;k\}}_{\ell',\ell}({\bs t})
\end{aligned}
\ee
where we used a combination of the tricks above.
Combining everything, we have
\be
\hat G^{\{-;k\}}_{\rm{int}}({\bs t}) = {\scr D}^{\{-;k\}}_\rm{int}({\bs t}) \, \hat G^\rm{prim}_\rm{int}({\bs t}),
\ee
where 
\be	\label{Jminsigma}
{\scr D}^{\{-;k\}}_{\rm{int}}({\bs t})
	= 
	- 2q^\fra_n
	\sum_{r = 0}^{k + R_\fra}
	A^{(\infty)}_{r} 
	A^{(0)}_{r} .
\ee

Thus we have seen that the functions (\ref{Gakint}), with excitations of NS chirals and anti-chirals, always factorize (on the covering) to a factor ${\scr D}^{\{a;k\}}_\rm{int}$ times a function without mode excitations. 
(Note that the same is not true for similar functions containing multiplet fields $\s^{\fra(r)}_{[n]}$ in place of the NS chirals; following the same steps above, it is not difficult to see that the obstruction to the factorization comes from the fact that
$( J^3_{\ell} \hat \s^{\fra(r)}_n) \neq 0$ 
for $\ell < R_\fra$, as shown in (\ref{VanisOfJ3modesMULTI}).)

\subsection{Four-point functions with twists $(n)$-$(2)$-$(2)$-$(n)$}		
\label{SectExample1nDesc}

If we want explicit formulas for the correlation functions, we need an explicit covering map, so that we can compute the $A$-coefficients. 
We will now specialize to the twist structure of the functions (\ref{Gakint}) with the interaction operator that we considered above.
Consider then the following class of base-space four-point function, with three points fixed by global conformal symmetry to be at $z = 0,1,\infty$, and the fourth point being at the cross-ratio $z=u$,
\be	\label{Gzexmvvvvqwkk}
G(u) = 
\Big\langle
\lord  J^{a\dagger}_{\frac{k}{n}} \s^{\fra\dagger}_{[n]}\rord (\infty) 
\ {\scr O}^\dagger_{[2]}(1)
\ {\scr O}_{[2]}(u)
\ \lord J^a_{-\frac{k}{n}} \s^{\fra}_{[n]} \rord (0)
\Big\rangle .
\ee
For the sake of concreteness, we will take ${\scr O}_{[2]}$ to be either the interaction operator or a NS chiral, but we emphasize that everything related to the covering map, e.g.~the $A$-coefficients, depends exclusively on the twists and would hold for any other choice of fields provided they have the same twist structure 
$(n)$-$(2)$-$(2)$-$(n)$.

\subsubsection{Covering map and $A$-coefficients}

The twist structure of the function (\ref{Gzexmvvvvqwkk}) corresponds to the covering map
\bsub	\label{coverm}
\be
z(t) = \left( \frac{t}{t_1}\right)^{n} 
	\left( \frac{t-t_0}{t - t_\infty} \right) 
	\left( \frac{t_1-t_\infty }{t_1-t_0} \right)
\ee
where 
\be	\label{covermtt}
t_0 = x-1 
\qquad
t_1 = \frac{(x-1) (x + n - 1)}{n +  x} 
\qquad
t_\infty = x- \frac{ x }{n+ x} .
\ee
\esub
The formulas (\ref{covermtt}) are fixed by requiring that $z(t)$ has a second-order ramification at $z = 0$ and $z = 1$, corresponding to the permutation cycles of length two in (\ref{Gzexmvvvvqwkk}).
The map is such that the points $z$ on base space where there are operator insertions in (\ref{Gzexmvvvvqwkk}) lift to points $t$ on covering space as
\be
z = [\infty, 1,u, 0 ] \quad \mapsfrom \quad t = [\infty, t_1, x, 0] .
\ee
In particular, the conformal cross-ratio on the base sphere, $u$, lifts to the point $t = x$, which is the only free parameter on the covering map: the parameters $t_1$, $t_0$, $t_\infty$ are all functions of $x$ only. 

The $A$-coefficients are given by Eqs.(\ref{AstaellDerccBell})-(\ref{AstaellDerccInf}) in terms of Bell polynomials which are computed in terms of the coefficients $c^{(*)}_m$ and $b_*$, defined by the expansions (\ref{ztgenercoveC})-(\ref{ztgenercoveInf}).
Given the covering map (\ref{coverm}), we can give an explicit formula for these coefficients. By expanding $z(t)$ around $t = 0$, 
\[
z(t) 
	=
	\frac{(t_1 - t_\infty)t_0}{(t_1 - t_0) t_\infty t_1^n} 
	\left(1 + \sum_{m=1}^\infty \frac{t_0 - t_\infty}{t_0 t_\infty^m} t^m \right) 	t^n ,
\]
comparing with (\ref{ztgenercoveC}) with $t_* = 0$, then substituting Eqs.(\ref{covermtt}) we find
\be	\label{b0andc0m}
\begin{aligned}
b_0 &= 
	x^{-1}
	(x-1)^{-n}
	(x + n)^{n + 1}
	(x + n - 1)^{-n} ,
\\
c^{(0)}_m 
	&= - n 
	x^{-m}
	(x-1)^{-1}
	(x+n)^{m-1}
	(x+n-1)^{-m}
\end{aligned}
\ee
From a similar expansion for $t = \infty$, we conclude that $c^{(\infty)}_m = (t_\infty - t_0) / t_\infty^{1-m}$, and, substituting Eqs.(\ref{covermtt}),
\be	\label{binfandcinfm}
\begin{aligned}
b_{\infty} &=
		(-1)^{n}
		(x-1)^{-(n+1)}
		(x + n)^{n}
		(x + n - 1)^{1-n}
\\
c^{(\infty)}_m 
	&= 
	n \,
	x^{m-1}
	(x+n)^{m-2}
	(x+n-1)^{-m+1}
\end{aligned}
\ee
Inserting (\ref{b0andc0m}) and (\ref{binfandcinfm}) into the Bell polynomials (\ref{AstaellDerccBell})-(\ref{AstaellDerccInf}), we can compute every $A$-coefficient as functions of the covering surface cross ratio $x$.
Note first that
\be	\label{ccoeffrtati}
\begin{aligned}
\frac{c_m^{(0)}}{c_{1}^{(0)}}  &= \gaO^{m-1} ,
	&\quad
\gaO &\equiv \frac{x+n}{x(x+n-1)} ,
	&\quad
c_1^{(0)} &= - \frac{n}{x(x-1)(x+n-1)} ,
\\
\frac{c_m^{(\infty)}}{c_{1}^{(\infty)}} 
	&
	= \gaInf^{m-1} ,
	&\quad
\gaInf &\equiv \frac{x(x+n-1)}{(x+n)}  ,
	&\quad
c_1^{(\infty)} &= \frac{n}{x+n},
\end{aligned}
\ee
so we can use the identity (\ref{IdentBellHomo}), with $b=c_1^{(*)} / \gamma_{*}$, $a=\gamma_{*}$ and $x_m=m!$ to write
\bsub	\label{A0Infn22n}
\begin{align}
\label{A0elln22nA}
A^{(0)}_\ell 
	&= 
	b_0^{-\frac{k}{n}}
	\sum_{m = 1}^\ell\mathcal{C}_{\ell,m}
	(k/n)_m
	\left[ \frac{n}{(x-1)(x+n)} \right]^m
	\left[ \frac{(x+n)}{x(x+n-1)} \right]^{\ell}
\\
\label{Ainfelln22nB}
A^{(\infty)}_\ell 
	&= 
	b_\infty^{\frac{k}{n}}
	\sum_{m = 1}^\ell\mathcal{C}_{\ell,m}
	(-k/n)_m
	\left[ \frac{-n}{x(x+n-1)} \right]^m
	\left[ \frac{x(x+n-1)}{(x+n)} \right]^{\ell}
\end{align}
\esub
where 
\be	\label{LahC}
{\cal C}_{\ell,m}\equiv \sum_{[j]}\frac{1}{j_1!j_2!\cdots j_{\ell-m+1}!} .
\ee
The sum is over $j_i$ configurations restricted by the conditions (\ref{BellPoluincomdefCond}).
These coefficients are related to the Lah numbers $L(\ell,m)$,
\be
	\mathcal{C}_{\ell,m}=\frac{1}{\ell!}L(\ell,m)=\frac{1}{m!}{\ell - 1 \choose m-1} .
\ee
Note that since $\ell>0$ the Bell Polynomial vanishes for $m=0$ hence the sums (\ref{A0Infn22n}) start at $m=1$.

\subsubsection{Sample functions}

The correlator (\ref{Gzexmvvvvqwkk}) lifts to
\be
\hat G(x) 
	= 
	\sum_{\ell, \ell' = - R_{\fra}}^{k} 
	A^{(\infty)}_{k-\ell'}
	A^{(0)}_{k-\ell}
	\Big\langle
	\lord  J^{a\dagger}_{\ell'} \hat \s^{\fra \dagger}_{n} \rord (\infty) 
	\; \hat {\scr O}^\dagger_{2}(t_1) 
	\; \hat {\scr O}_{2}(x) \;
	\lord J^a_{- \ell} \hat \s^\fra_n \rord (0)
	\Big\rangle .
\ee
We can now give examples of explicit formulas for these functions. 
We will restrict ourselves to the cases where $\hat G(x)$ can be expressed in terms of a correlator without excitations, i.e.
\be	\label{hatGscrDhatGprim}
\hat G (x) = {\scr D}(x) \hat G_\rm{prim} (x) ,
\ee
where
\be
\hat G_\rm{prim} (x) 
	=
	\Big\langle
	\hat \s^{\fra \dagger}_{n} (\infty) 
	\; \hat {\scr O}^\dagger_{2}(t_1) 
	\; \hat {\scr O}_{2}(x) \;
	\hat \s^\fra_n (0)
	\Big\rangle .
\ee
As we have seen, this factorization happens for functions with the deformation operator, and also for the special case of functions with NS chirals and anti-chirals excited by $J^3$ modes, given in Eq.(\ref{GNSalgeJ3BANTI}), (\ref{J3sigmaINT}) and (\ref{Jminsigma}).

\subsubsection*{Functions with NS chirals and anti-chirals}

The only configuration with zero charge compatible with both (\ref{Gzexmvvvvqwkk}) and (\ref{4ptFunctkknn}), and including chirals and anti-chirals is 
\be	\label{4ptFuncNS3kn22n}
G_\NS^{\{3;k\}}(u) 
	=
\Big\langle
\lord  J^{3}_{\frac{k}{n}} \s^{\fra \dagger}_{[n]}\rord (\infty) 
\; \s^{\frb \dagger}_{[2]}(1)
\; \s^{\frb}_{[2]}(u)
\lord J^3_{-\frac{k}{n}} \s^{\fra}_{[n]} \rord (0)
\Big\rangle .
\ee
It lifts to
\be
\hat G_\NS^{\{3;k\}} (x) 
	= 
	{\scr D}_\NS^{\{3;k\}}(x)
	\
	\Big\langle
	\lord  \hat \s^{\fra \dagger}_{n}\rord (\infty) 
	\; \hat \s^{\frb \dagger}_{2}(t_1)
	\; \hat \s^{\frb}_{2}(x)
	\hat \s^{\fra}_{n} (0)
\Big\rangle ,
\ee
where 
\be
\begin{aligned}
&{\scr D}_\NS^{\{3;k\}}(x)
	=
\\
&	\sum_{\ell , \ell' = 0}^{k} 
	A^{(\infty)}_{k'-\ell'}
	A^{(0)}_{k-\ell}
	\Bigg[
	- \frac{\ell}{2} \delta_{\ell',\ell}
	- (q^\fra_n)^2 \delta_{\ell', 0} \delta_{\ell, 0}
	+
	\theta(\ell) 
	\big(
	\delta_{\ell',0} \, q^\fra_n 
	+
	q^{\frb}_{2} t_2^{\ell'}
	\big)
	\big(
	q^{\frb}_{2} \, t_2^{-\ell} - q^{\fra}_{n}
	\big)
\\
&\qquad\qquad\qquad	
	+ q^{\frb}_{2}
	\Big(
	 q^\fra_n t_1^{\ell'} \ \delta_{\ell,0}
	+ \theta(\ell) \ t_1^{\ell'} \ \big( q^{\frb}_{2} \, x^{-\ell} - q^{\fra}_{n} \big)
	-
	q^{\frb}_{2} x^{\ell'} t_1^{- \ell}
	-
	q^{\frb}_{2} t_1^{- \ell + \ell'} 
	\Big)
	\Bigg] .
\end{aligned}
\ee
The expression in square brackets was computed in (\ref{GNSalgeJ3BANTI}).
The explicit formulas for ${\scr D}$ are not very illuminating, but it is worthwhile to quote the simplest ones. The simplest case is the lowest excitation with $k=1$, 
\be	\label{scrD31NS}
{\scr D}^{\{3;1\}}_\NS(x)
	= 
	- \left[
	\frac{x ( x+n-1)}{(x+n)(x-1)} 
	\right]^{\frac{1}{n}}
	\Bigg[
	\frac12
	+
	\frac{\big( q^\fra_n + q^\frb_2(2x+n-1) \big)^2}
		{x (x-1)(x+n)(x+n-1)}
	\Bigg] .
\ee
For $k = 2$ we already have a much more cumbersome formula:
{\footnotesize
\be
\begin{aligned}
&{\scr D}^{\{3;2\}}_\NS(x)
= 
\\
&	\left[
	\frac{x ( x+n-1)}{(x+n)(x-1)} 
	\right]^{\frac{2}{n}}
	\Bigg[
	1 
	+
	\frac{2}{x (x-1)(x+n)(x+n-1)}
\\
&	+	
	\frac{
		(q^\frb_2)^2 (2x + n-1)^2 [ 2 x (x-1) + n(2x -1) + 3]^2
		}
		{x^2 (x-1)^2 (x+n)^2 (x+n-1)^2}
\\
&	+
	\frac{
	n^2 (q^\fra_n)^2 \big[ n (1-2 x)^2 + 12 x^3-12 x^2+13 x-5 \big]
	+ 2 n^3 q^\frb_2 q^\fra_n \big[ n (1-2 x)^2 +  20 x^3-28 x^2+23 x-7 \big]
		}{(x-1)^2 x^2 (n+x-1)^3 (n+x)^2}
\\
&	+ 
	\frac{
	2 (q^\fra_n)^2 
	\big[ 
	2 ( x^5-x^4+2 x^3-2 x^2+2 x-1 )
	+ n (6 x^4-6 x^3+10 x^2-7 x+4 )
	\big]
	}{
	(x-1)^2 x^2 (n+x-1)^3 (n+x)^2
	}
\\
&	+
	\frac{ 
	2 n^2  q^\frb_2 q^\fra_n \big[ 36 x^4-64 x^3+82 x^2-51 x+17 \big]
	+ 2 n q^\frb_2 q^\fra_n \big[ 28 x^5-60 x^4+100 x^3-92 x^2+57 x-17 \big]
	}{(x-1)^2 x^2 (n+x-1)^3 (n+x)^2}
\\
&	+ \frac{ 2 q^\frb_2 q^\fra_n \big[
	 8 x^6-20 x^5+40 x^4-48 x^3+42 x^2-25 x+6
	\big]
		}{(x-1)^2 x^2 (n+x-1)^3 (n+x)^2}
	\Bigg] .
\end{aligned}
\ee
}%
In general, we will have even longer rational functions of $x$. The square brackets to the power $k/n$ at the beginning of the formulas correspond to the factors of $(b_0/b_\infty)^{k/n}$ that come from the liftings.

\subsubsection*{Functions with the deformation operator}

With the interaction operator, the function
\be	\label{4ptFuncNS3kn22n}
G_\rm{int}^{\{a;k;\fra\}}(u) 
	=
\Big\langle
\lord  J^{a\dagger}_{\frac{k}{n}} \s^{\fra \dagger}_{[n]}\rord (\infty) 
\; \Oint(1)
\; \Oint(u)
\lord J^a_{-\frac{k}{n}} \s^{\fra}_{[n]} \rord (0)
\Big\rangle 
\ee
lifts to
\be
\hat G_\rm{int}^{\{3;k;\fra\}} (x) 
	= 
	{\scr D}_\rm{int}^{\{a;k\}}(x)
	\
	\Big\langle
	 \hat \s^{\fra \dagger}_{n}(\infty)  
	\; \Oincov(t_1)
	\; \Oincov(x)
	\; \hat \s^{\fra}_{n} (0)
\Big\rangle ,
\ee
where ${\scr D}_\rm{int}^{\{a;k\}}$ is given by (\ref{J3sigmaINT}) and (\ref{Jminsigma}). 
Note that, in both cases, the sum over $A$-coefficients is ``diagonal'', and we can give an explicit formula for the product
\be	\label{A0AInfkk}
\begin{aligned}
&A_{\ell}^{(0)}A_{\ell}^{(\infty)} 
	= 
\\
&
	\left[	\frac{x ( x+n-1)}{(x+n)(x-1)} \right]^{\frac{\ell}{n}}
	\sum_{p,q=1}^{\ell}
	{\cal C}_{\ell,p} {\cal C}_{\ell,q}
	(\ell/n)_{p} 
	(-\ell/n)_{q}
	\left[ \frac{n}{(x-1)(x+n)} \right]^p
	\left[ \frac{-n}{x(x+n-1)} \right]^q 
\end{aligned}
\ee
where the ${\cal C}_{\ell,p}$ are given by (\ref{LahC}). 
This expression gives the explicit dependence of ${\scr D}^{\{a;k\}}_\rm{int}$ with $x$, which allows us to track the OPE channels of the four-point function with relative ease, as we will show in Sect.\ref{SectFunctsonBase}.
The explicit formulas are, however, again very cumbersome except for the simplest cases.

The lowest excitation with $k =1$ for $J^3$ modes is
\be		\label{scrDint31}
{\scr D}^{\{3;1;\fra\}}_{\rm{int}}(x) 
	=
	-
	\left[	\frac{x ( x+n-1)}{(x+n)(x-1)} \right]^{\frac{1}{n}}
	\left[
		\frac{1}{2}
		+ \frac{(q^\fra_n)^2}{x(x-1)(x+n)(x+n-1)} 
	\right] .
\ee
For $k =2$,
{\footnotesize
\be
\begin{aligned}
&{\scr D}^{\{3;2;\fra\}}_{\rm{int}}(x)  =
\\
&-	\left[	\frac{x ( x+n-1)}{(x+n)(x-1)} \right]^{\frac{2}{n}}
	\Bigg[
	1
	+ \frac{2}{x(x-1)(x+n)(x+n-1)}
\\
&
	+ \frac{
	2(q^\fra_n)^2 
	\big[ x^2 + (n-1)x + 1 - \frac12 n \big]
	\big[
	2 (x^3 + x - 1)
	+ n (4x^2 - x + 3)
	+ n^2 (2x -1)
	\big]
	}{x^2 (x-1)^2 (x+n)^2 (x+n-1)^3} 
	\Bigg]
\end{aligned}
\ee
}%
In general, we will have long rational functions of $x$ (times the factors $(b_0/b_\infty)^{k/n}$).
For excitations by $J^-$ we have the further complication of the sum limits increasing with $n$ as well as with $k$. The simplest non-trivial case is $k = 1$, $\fra = -1$ and $n = 2$,
\be 	\label{scrDintpm1}
{\scr D}^{\{-,-;-\}}_{\rm{int}}(x) 
	=
	- \left[	\frac{x ( x+1)}{(x+2)(x-1)} \right]^{\frac{1}{2}}
	\left[
	1
	+ \frac{1}{x(x-1)(x+2)(x+1)} 
	\right] 
\quad 
	(n=2)
	.
\ee
As with the correlators involving NS chirals, for any other values of the parameters $k, n$, etc., the functions become increasingly less illuminating.

\subsection{Four-point functions with twists $(n_1)(n_2)$-$(2)$-$(2)$-$(n_1)(n_2)$} \label{Sect4ptBoubleCycle}

We want to consider generalizations of the $(n)$-$(2)$-$(2)$-$(n)$ four-point functions studied in Sect.\ref{SectExample1nDesc} by replacing the single cycles $(n)$ by double-cycled states/fields, that is, states $\ket{{\scr O}}_g$ with a permutation $g = (n_1)(n_2)$, with two disjoint non-trivial cycles of lengths $n_1,n_2 > 1$. These states are tensor products of two ``string components'', i.e.~two single-cycle states, of the form
\be	\label{Staten1n2}
	\ket{{\scr O}^1}_{n_1} \otimes \ket{{\scr O}^2}_{n_2} \otimes \left( \ket{\varnothing}_1\right)^{N-n_1-n_2} 
	\cong
	\big[ {\scr O}^1_{[n_1]} {\scr O}^2_{[n_2]} \big] (0) .
\ee
For definiteness, we will take each string component to be an R-current excitation of an NS chiral field/state,
\be
	J^{a_1}_{-\frac{k_1}{n_1}} \ket{\s^{{\frak a}_1}}_{n_1} 
	\otimes 
	J^{a_2}_{-\frac{k_2}{n_2}} \ket{\s^{{\frak a}_2}}_{n_2} 
	\otimes \left( \ket{\varnothing}_1\right)^{N-n_1-n_2} 
	\cong
	\Big[ 
	\lord J^{a_1}_{-\frac{k_1}{n_1}} \s^{\frak a_1}_{[n_1]} \rord
	\lord J^{a_2}_{-\frac{k_2}{n_2}} \s^{\frak a_2}_{[n_2]} \rord
	\Big] (0) ,
\ee
and we consider the four-point function
\be	\label{GNdoublCycl0}
G (u)
	= 
	\Big\langle
	\Big[ 
	\lord J^{a_1\dagger}_{\frac{k_1}{n_1}} \s^{\frak a_1\dagger}_{[n_1]}  \rord
	\lord J^{a_2\dagger}_{\frac{k_2}{n_2}} \s^{\frak a_2\dagger}_{[n_2]} \rord
	\Big] (\infty)
	{\scr O}^\dagger_{[2]}(u) 
	{\scr O}_{[2]}(1)
	\Big[ 
	\lord J^{a_1}_{-\frac{k_1}{n_1}} \s^{\frak a_1}_{[n_1]} \rord 
	\lord J^{a_2}_{-\frac{k_2}{n_2}} \s^{\frak a_2}_{[n_2]} \rord
	\Big] (0)
	\Big\rangle 
\ee
where ${\scr O}_{[2]}$ can be the interaction field or an NS field, like in (\ref{Gzexmvvvvqwkk}).
In the large-$N$ limit, after a factorization, the leading contribution to four-point functions of the type
$\bra{\Psi} {\scr O}_{[2]}^\dagger(u) {\scr O}_{[2]}(1) \ket{\Psi} ,$
where $\ket{\Psi}$ is a \emph{generic,} multi-cycle state with an arbitrary permutation, are functions where $\ket{\Psi}$ is of the form (\ref{Staten1n2}) \cite{Lima:2022cnq}.
Our interest here is in illustrating some interesting properties of the lifting of modes in maps with multiple cycles. 
The covering map that gives a genus-zero covering surface is a generalization of (\ref{coverm}), the points $t_0, t_\infty$ being now ramification points corresponding to the non-trivial cycles of length $n_2$,
\bsub	\label{coverm2}
\be
z(t) = \left( \frac{t}{t_1}\right)^{n_1} 
	\left( \frac{t-t_0}{t - t_\infty} \right)^{n_2} 
	\left( \frac{t_1-t_\infty }{t_1-t_0} \right)^{n_2} ,
\ee
with
\be	\label{tim}
\begin{split}
t_0 = x-1 ,
\qquad
t_1 = \frac{(x-1) (n_1+ n_2 x- n_2)}{n_1 + n_2 x} ,
\qquad
t_\infty = x- \frac{n_2 x }{n_1+n_2 x} .
\end{split}
\ee
\esub
Cf.~\cite{Lima:2022cnq}. The four $A$-coefficients can be computed as before, expanding the map around the four ramification points where there are current insertions.

\subsubsection{Lifting of a double cycle}

In general, the covering map $t \mapsto z(t)$ for correlator with a double-cycle field
$\big[ {\scr O}^1_{[n_1]} {\scr O}^2_{[n_2]} \big]$
is such that the ramification points corresponding to the cycles $(n_1)$ and $(n_2)$ map to the same point on the base.
For a correlation function like (\ref{GNdoublCycl0}), 
{\footnotesize
\be	\label{Mappingttinfo}
\begin{tikzcd}
 t=\infty \ar[dr] & & t=t_\infty \ar[dl]  \\
 & z=\infty & \\
\end{tikzcd}
\quad \
\begin{tikzcd}
 t = t_1 \ar[d] &  \\
 z = 1 & \\
\end{tikzcd}
\!\!\!\!
\begin{tikzcd}
t=x \ar[d] & \\
z=u & \\
\end{tikzcd}
\begin{tikzcd}
 t=0 \ar[dr] & & t=t_0 \ar[dl]  \\
 & z=0 & \\
\end{tikzcd}
\ee
}%
Thus the lifted correlator is a six-point function. 
The contour integrals that make the mode excitations of each cycle are around different points, and give independent sums.
Let us illustrate this for the operators at infinity.

The point $t_\infty$ is slightly different from the ramification points we have considered before. It is a finite point on the covering which maps to infinity at the base. Around such a point, the covering map with a ramification of order $n$ behaves as 
\be
z(t) = 
	b_{t_\infty} (t-t_\infty)^{-n} 
		\Big[1 + c^{(t_\infty)}_{1} (t-t_\infty) + c^{(t_\infty)}_{2}(t-t_\infty)^2 + \cdots \Big]
\ee
Note the negative power, in comparison with (\ref{ztgenercoveC}) and (\ref{ztgenercoveInf}).
Therefore
\be	\label{ztoknattinft}
[z(t)]^{\frac{k}{n}} 
	=
	(t- t_\infty)^{-k}
	\sum_{\ell=0}^\infty A_{\ell}^{(t_\infty)} (t-t_\infty)^\ell ,
\ee
where
\be
\begin{aligned}
A^{(t_\infty)}_\ell 
&	= 
	\left.
	\frac1{\ell!} \frac{d^\ell}{dt^\ell} 
	\left[ \frac{z(t)}{(t-t_\infty)^n} \right]^{\frac{k}{n}} 
	\right|_{t=t_\infty} 
\\
&	=
	\frac{b_{t_\infty}^{\frac{k}{n}}}{\ell!} \frac{d^\ell}{dt^\ell} 
	\left[ 
	1 + c^{(t_\infty)}_{1} (t-t_\infty) + c^{(t_\infty)}_{2}(t-t_\infty)^2 + \cdots
	\right]^{\frac{k}{n}} 
	\Bigg|_{t=t_\infty} 
\\
&	= 
	\frac{b_{t_\infty}^{\frac{k}{n}}}{\ell!}
	\sum_{m = 0}^\ell
	(-)^m
	(-k/n)_m 
	\Bell_{\ell, m} 
		\big[ c^{(t_\infty)}_1 , 
			2! c^{(t_\infty)}_2, \cdots, 
			(\ell - m +1)! c^{(t_\infty)}_{\ell-m+1}
		\big]
\end{aligned}
\ee
Because of the negative sign in the powers of $(t-t_\infty)$ in (\ref{ztoknattinft}), it follows that a field at infinity on the base lifts by (\ref{coverm2}) to
\be
\begin{aligned}
&\Big[ 
\lord J^{a_1}_{\frac{k_1}{n_1}} {\scr O}^1_{[n_1]}  \rord
\lord J^{a_2}_{\frac{k_2}{n_2}} {\scr O}^2_{[n_2]} \rord
\Big] (\infty)
\\
&\qquad\qquad\upmapsto
\\
&	\Big(
	\oint_{\infty} \frac{d t}{2\pi i}
	[z(t)]^{\frac{k_1}{n_1}} 
	J^{a_1} (t) \hat {\scr O}^1_{n_1}(\infty) 
	\Big)
	\left(
	\oint_{t_\infty} \frac{d t}{2\pi i}
	[z(t)]^{\frac{k_2}{n_2}} 
	J^{a_2} (t) \hat {\scr O}^2_{n_2}(t_\infty) 
	\right)
\\
&=
	\oint_{\infty} \frac{d t'}{2\pi i}
	t'^{k_1}
	 \left[ \sum_{\ell_1=0}^\infty A^{(\infty)}_{\ell'} t'^{-\ell_1} \right]
	J^{a_1} (t_1) \hat {\scr O}^1_{n_1}(\infty) 
\\
&\qquad
	\times
	\oint_{t_\infty} \frac{d t}{2\pi i}
	(t - t_\infty)^{-k_2}
	 \left[ \sum_{\ell_2=0}^\infty A^{(t_\infty)}_{\ell_2} (t-t_\infty)^{\ell_2} \right]
	J^{a_2} (t) \hat {\scr O}^2_{n_2}(t_\infty) 
\\
&=
	 \left[ 
	 \sum_{\ell'=0}^\infty A^{(\infty)}_{\ell_1} 
	\lord  J^{a_1}_{k_1 - \ell_1} \hat {\scr O}^1_{n_1} \rord (\infty)
	 \right]
	 \left[ 
	 \sum_{\ell=0}^\infty A^{(t_\infty)}_{\ell_2} 
	\lord  J^{a_2}_{-k_2 + \ell_2} \hat {\scr O}^2_{n_2} \rord  (t_\infty)
	 \right]
\end{aligned}
\ee
Note that $k_1$ and $k_2$ ``lift with opposite signs''. This is consistent with the fact that $(J^a_k {\scr O})(\infty) = (J^a_{-k} {\scr O}) (0)$, so if $(J^a_k {\scr O}) (\infty) \neq 0$ on the base, it should lift, schematically, to $(J^a_{-k} \hat {\scr O}) (t_\infty)$ on the covering.

The lifting of double-cycle fields at points $z \neq \infty$ on the base sphere is similar but simpler, without the caveat of changing the sign of the lifted modes.
Therefore functions with the twist structure of (\ref{GNdoublCycl0}) lift to 
\bsub	\label{GtsumellDoubleNS}
\be
\hat G(x)
	= 
	\sum_{\ell_1, \ell_1' = -R_{\frak a_1}}^{k_1} 
	\sum_{\ell_2, \ell_2' = - R_{\frak a_2}}^{k_2}
	A^{(\infty)}_{k_1'-\ell_1'}
	A^{(t_\infty)}_{k_2'-\ell_2'}
	A^{(0)}_{k_1-\ell_1}
	A^{(t_0)}_{k_2-\ell_2}
	\hat G_{\ell_1',\ell_1; \ell_2',\ell_2}^{\{a_i;{\fra}_i\}}(x)
\ee
which is a superposition of the \emph{six}-point functions
\be	\label{GtsumellDoubleNS2}
\begin{aligned}
&	\hat G_{\ell_1',\ell_1; \ell_2',\ell_2}^{\{a_i;{\frak a}_i\}}(x)
	=
\\
&
	\Big\langle
	\lord J^{a_1\dagger}_{\ell'_1} \hat \s^{\frak a_1\dagger}_{n_1} \rord  (\infty) \
	\lord J^{a_2\dagger}_{-\ell'_2} \hat \s^{\frak a_2\dagger}_{n_2} \rord (t_\infty) \
	\hat {\scr O}_2^{\dagger}(x) \
	\hat {\scr O}_{2}(t_1) \
	\lord J^{a_1}_{-\ell_1} \hat \s^{\frak a_1}_{n_1} \rord  (0) \
	\lord J^{a_2}_{-\ell_2} \hat \s^{\frak a_2}_{n_2} \rord  (t_0)
	\Big\rangle 
\end{aligned}
\ee
\esub
We can now proceed as before, using contour deformations to move the $J^{a\dagger}$ modes to the other fields. (Note that this can be done since the covering surface defined by (\ref{coverm2}) is a sphere.)

\subsubsection{Functions with the deformation operator}

Consider the functions with $\Oint$ and $J^3$ excitations only. In this case we have $R_{\fra_1} = R_{\fra_2} = 0$ hence all $\ell_i \geq 0$.
We can use the Ward identities to eliminate $J^3_{\ell_1'}$,
{\footnotesize
\be	\label{GtsumellDoubleNS2J33}
\begin{aligned}
&	\hat G_{\ell_1',\ell_1; \ell_2',\ell_2}^{\{3; \fra_i\}}(x)
	=
	\Big\langle
	J^{3}_{\ell'_1} \hat \s^{\frak a_1\dagger}_{n_1}  (\infty) \
	J^{3}_{-\ell'_2} \hat \s^{\frak a_2\dagger}_{n_2} (t_\infty) \
	\Oincov(x) \
	\Oincov (t_1) \
	J^{3}_{-\ell_2} \hat \s^{\frak a_2}_{n_2} (t_0) \
	J^{3}_{-\ell_1} \hat \s^{\frak a_1}_{n_1}  (0) 
	\Big\rangle 
\\
&	=
	- \Big( q^{\fra_1}_{n_1} \delta_{\ell'_1,0} - \frac{\ell_1}{2} \delta_{\ell_1,\ell'_1} \Big)
	\Big\langle
	\hat \s^{\frak a_1\dagger}_{n_1}  (\infty) \
	J^{3}_{-\ell'_2} \hat \s^{\frak a_2\dagger}_{n_2} (t_\infty) \
	\Oincov(x) \
	\Oincov (t_1) \
	J^{3}_{-\ell_2} \hat \s^{\frak a_2}_{n_2} (t_0) \
	\hat \s^{\frak a_1}_{n_1}  (0) 
	\Big\rangle 
\\
&
	-
	\Bigg[
		\theta(\ell'_1 - \ell'_2) {\ell'_1 \choose \ell'_2} \frac{\ell'_2}{2} t_\infty^{\ell'_1 - \ell'_2}
		- q^{\fra_2}_{n_2} t_\infty^{\ell_1'} 
	\Bigg]
	\Big\langle
	\hat \s^{\frak a_1\dagger}_{n_1}  (\infty) 
	J^{3}_{-\ell'_2} \hat \s^{\frak a_2\dagger}_{n_2} (t_\infty) 
	\Oincov(x) 
	\Oincov (t_1) 
	\hat \s^{\frak a_2}_{n_2} (t_0) 
	J^{3}_{-\ell_1} \hat \s^{\frak a_1}_{n_1}  (0) 
	\Big\rangle
\\
&
	-
	\Bigg[
		\theta(\ell'_1 - \ell_2) {\ell'_1 \choose \ell_2} \frac{\ell_2}{2} t_\infty^{\ell'_1 - \ell_2}
		+ q^{\fra_2}_{n_2} t_\infty^{\ell_1'} 
	\Bigg]
	\Big\langle
	\hat \s^{\frak a_1\dagger}_{n_1}  (\infty) 
	\hat \s^{\frak a_2\dagger}_{n_2} (t_\infty) 
	\Oincov(x) 
	\Oincov (t_1) 
	J^{3}_{- \ell_2} \hat \s^{\frak a_2}_{n_2} (t_0) 
	J^{3}_{-\ell_1} \hat \s^{\frak a_1}_{n_1}  (0) 
	\Big\rangle
\end{aligned}
\ee
}%
where $\theta(\xi) = 1$ for $\xi > 0$ and zero otherwise.
(Here we omit the parentheses in the excited fields for brevity.)
Each remaining function has two $J$ insertions. We can get rid of them as well.
For example, the correlator in the second line can be written as
{\small
\be	\label{Gtdad333das}
\begin{aligned}
&	\Big\langle
	\hat \s^{\frak a_1\dagger}_{n_1}  (\infty) \
	J^{3}_{-\ell'_2} \hat \s^{\frak a_2\dagger}_{n_2} (t_\infty) \
	\Oincov(x) \
	\Oincov (t_1) \
	J^{3}_{-\ell_2} \hat \s^{\frak a_2}_{n_2} (t_0) \
	\hat \s^{\frak a_1}_{n_1}  (0) 
	\Big\rangle 
\\
&	=
	\Bigg[
	q^{\fra_1}_{n_1}  ( \delta_{\ell'_2,0} - t_0^{\ell_2} )
		\big[ q^{\fra_1}_{n_1} ( \delta_{\ell_2,0} - t_0^{\ell_2}) + q^{\fra_2}_{n_2} (t_0 - t_\infty)^{\ell_2} \big]
\\
&\qquad\qquad\qquad
	- q^{\fra_2}_{n_2} (t_0 - t_\infty)^{\ell_2'}
	+ (-)^{\ell_2} \frac{\ell_2}{2} { \ell_2 + \ell'_2 -1 \choose \ell_2} (t_0 - t_\infty)^{\ell_2+\ell'_2}
	\Bigg]
\\
&\qquad\qquad\qquad\qquad\qquad
	\times
	\Big\langle
	\hat \s^{\frak a_1\dagger}_{n_1}  (\infty) \
	\hat \s^{\frak a_2\dagger}_{n_2} (t_\infty) \
	\Oincov(x) \
	\Oincov (t_1) \
	\s^{\frak a_2}_{n_2} (t_0) \
	\hat \s^{\frak a_1}_{n_1}  (0) 
	\Big\rangle 
\end{aligned}
\ee
}%
and similar expressions hold for the other correlators. 
Thus in the end (\ref{GtsumellDoubleNS2J33}) can be fully reduced to the form 
\be
\hat G_{\ell_1',\ell_1; \ell_2',\ell_2}^{\{3;\fra_i\}}(x)
	= {\scr D}_{\ell_1',\ell_1; \ell_2',\ell_2}^{\{3;\fra_i\}}(x) \, \hat G_\rm{int}^\rm{prim}(x)
\ee
where
\be
\hat G_\rm{int}^\rm{prim}(x) = 
	\Big\langle
	\hat \s^{\frak a_1\dagger}_{n_1}  (\infty) \
	\hat \s^{\frak a_2\dagger}_{n_2} (t_\infty) \
	\Oincov(x) \
	\Oincov (t_1) \
	\s^{\frak a_2}_{n_2} (t_0) \
	\hat \s^{\frak a_1}_{n_1}  (0) 
	\Big\rangle 
\ee
is the covering-surface correlator without excitations.

For functions with $J^\pm$ excitations this factorization does not happen.
For example, consider
{\footnotesize
\be	\label{GtsumellDoubleNS2Jpm}
\begin{aligned}
&	\hat G_{\ell_1',\ell_1; \ell_2',\ell_2}^{\{-,-; \fra_i\}}(x)
	=
	\Big\langle
	J^{+}_{\ell'_1} \hat \s^{\frak a_1\dagger}_{n_1}  (\infty) \
	J^{+}_{-\ell'_2} \hat \s^{\frak a_2\dagger}_{n_2} (t_\infty) \
	\Oincov(x) \
	\Oincov (t_1) \
	J^{-}_{-\ell_2} \hat \s^{\frak a_2}_{n_2} (t_0) \
	J^{-}_{-\ell_1} \hat \s^{\frak a_1}_{n_1}  (0) 
	\Big\rangle 
\end{aligned}
\ee
}%
Using a Ward identity to move $J^+_{\ell'_1}$ from infinity to the other points will produce several functions, one of which is
\be
	\Big\langle
	\hat \s^{\frak a_1\dagger}_{n_1}  (\infty) \
	J^{+}_{s} J^{+}_{-\ell'_2} \hat \s^{\frak a_2\dagger}_{n_2} (t_\infty) \
	\Oincov(x) \
	\Oincov (t_1) \
	J^-_{- \ell_2}\hat \s^{\frak a_2}_{n_2} (t_0) \
	J^-_{- \ell_1} \hat \s^{\frak a_1}_{n_1}  (0) 
	\Big\rangle 
\ee
with $s \geq 0$.
A little thought shows that we cannot get rid of all current insertions. The trouble is, essentially, the same as that in Sect.\ref{SectFunctNSchiral}, related to the fact that there are non-vanishing positive-mode excitations of $\hat \s^\fra_n$ by $J^-$ and of $\hat \s^{\fra\dagger}_n$ by $J^+$.

\subsection{Functions on the base sphere; fusion rules}	\label{SectFunctsonBase}

Although the explicit formulas for the exact functions on the covering can become cumbersome, we will now show that we can obtain important information about the coincidence limits of operators inside the correlators. 
We are interested in the functions of the form (\ref{hatGscrDhatGprim}), 
\be	\label{hatGscrDhatGprim1}
\hat G^{\{a;k;\fra\}} (x) 
	= {\scr D}^{\{a;k;\fra\}} (x) \hat G_\rm{prim} (x) .
\ee
These functions, expressed in terms of the covering surface parameter $x$, are not the final answer. The \emph{base}-space functions $G(u)$, which are the final, physical result that we want, are obtained from  $\hat G(x)$ by inverting the covering map to find $u(x)$. By construction, the covering map is multivalued, and there are 
\be
{\bf H} = 2n
\ee
different inverses $x_{\frh}(u)$, $\frh = 1, \cdots, {\bf H}$, where ${\bf H}$ is the Hurwitz number counting the topologically different ramified coverings of the base sphere with the ramification data prescribed by the twists.
The complete correlator on the base is a sum of these `Hurwitz blocks,'
\be	\label{HurwExpanGkk}
\begin{aligned}
G^{\{a;k;\fra\}} (u) 
	= {\scr C}(N) \sum_{{\frh} = 1}^{\bf H} {\scr D}^{\{a;k;\fra\}} (x_{\frh}(u)) \hat G_\rm{prim} (x_{\frh}(u)) 
\end{aligned}
\ee
where ${\scr C}(N)$ is a symmetry factor.
We will consider now functions with twist structure $(n)$-$(2)$-$(2)$-$(n)$, in which case
${\scr C}(N) = 2n(N-2)! / N!$. A similar analysis holds for the double-cycle functions as well, see \cite{Lima:2022cnq}.

We want to explore the OPE channels by making $u$ go to the insertion points $z_i = 0,1,\infty$. 
In the covering map (\ref{coverm}), after writing $t_0$, $t_1$ and $t_\infty$ in terms of $x$, we find the following relation between the base-space cross-ratio $u = z(x)$ and its pre-image:
\be		\label{uxm}
u(x) = \Bigg( \frac{x+ n}{x-1} \Bigg)^{n+1} 
	\Bigg( \frac{x}{x + n - 1 } \Bigg)^{n-1}  .
\ee
The $2n$ functions $x_\frh(u)$ are obtained by inverting this polynomial relation.
Since we are interested in the OPE limits, we can compute the inverses asymptotically for $u$ near $z = 0$ and $z =1$. 

Each $x_\frh(u)$ corresponds to a class of combinations between the permutation cycles of the field at $u$, with the cycles of the other operator at $z_i$.
When $u \to 1$, there are two different solutions for (\ref{uxm}),
\bsub	\label{xaxp4ns}
\begin{align}
x &\to \infty , 
	& \quad x_{\frak1}(u) &\approx -{4n\over 1-u}+ \frac{3n+1}{2}+ \cdots
\\
 x &\to \tfrac{1-n}{2} , 
 	&\quad 
	x_{\frak2}(u) &\approx 
		\begin{aligned}[t]
		\frac{1-n}{2} 
				&+ \tfrac{3^{1/3} (n^2 -1)^{2/3}}{4n^{1/3}} (1 - u)^{\frac{1}{3}} 
				- \tfrac{3(n^2 +1)}{40n} (1-u) 
				\\
				&+\tfrac{ (n^2 -1)^{2/3}}{8 \cdot 3^{2/3} \cdot n^{1/3}} (1 -u)^{\frac{4}{3}} 
				+ \cdots
		\end{aligned}
\end{align}
\esub
This limit corresponds to the fusion
\be
{\scr O}_{(2)_u} \times {\scr O}_{(2)_1}^\dagger
\ee
inside the function (\ref{Gzexmvvvvqwkk}).
We have labeled the cycles temporarily according to their position on the base sphere. 
It can be shown 
\cite{Pakman:2009mi,Pakman:2009zz}
that if the covering surface is a sphere the permutations $(2)_u$ and $(2)_1$ are not disjoint, so there are two possibilities for their composition:
\be	\label{Channid3}
(2)_u (2)_1 = \rm{id} 
\qquad \text{or} \qquad
(2)_u (2)_1 = (3) .
\ee 
Which one is realized depends on the conjugacy classes of $(2)_u$ and $(2)_1$, and all conjugacy classes contribute to (\ref{HurwExpanGkk}).
Thus each of the combinations gives a different channel for the Hurwitz block expansion. 
The multiplicity of the solution 
$x_{\frak 1}(1) = \infty$ is 1, which means that the covering surface has only one sheet at $t = \infty$, hence this is an unramified point, hence it corresponds to an untwisted operator: this is the identity channel. 
The multiplicity of $x_{\frak 2}(1) = \frac{1-n}{2}$ is 3, hence three sheets meet at $t = x_{\frak 2}(1)$, and this corresponds to an operator with twist 3: this is the second possibility in (\ref{Channid3}). 

The more interesting fusion happens in the limit $u \to 0$, where we have
\be	\label{Channu0}
{\scr O}_{(2)_u} \times \ \lord J^a_{-\frac{k}{n}} \s^\fra_{(n)_0}\rord 
\ee 
For functions corresponding to a genus zero covering, we have again two possibilities for the cycles' composition \cite{Pakman:2009mi,Pakman:2009zz}:
\be	\label{Channid3}
(2)_u (n)_0 = (n\pm1) .
\ee 
The asymptotic solutions for (\ref{uxm}) are now
\bsub	\label{xaxp4ns}
\begin{align}
(2)_u (n)_0 &= (n-1):
&\quad
x &\to 0 , 
	& \quad x_{\frak3}(u) &\approx \left( \frac{(1-n)^{n-1}}{n^{n+1}} u\right)^{\frac{1}{n-1}} 
\\
(2)_u (n)_0 &= (n+1):
&\quad
 x &\to -n , 
 	&\quad 
	x_{\frak4}(u) &\approx - n + \left( \frac{(-n-1)^{n+1} }{n^{n-1}} u \right)^{\frac{1}{n+1}}
\end{align}
\esub
The multiplicities indicates that $x_{\frak 3}$ gives the channel where $(2)_u (n)_0 = (n-1)$ and $x_{\frak 4}$ the channel $(2)_u (n)_0 = (n+1)$.

We will assume that we know $\hat G_\rm{prim}(x)$. These have been computed in the literature, for instance in \cite{Lima:2022cnq}. Specifically, let us assume that $\hat G_\rm{prim}(x)$  has the asymptotics
\be	\label{hatGprimAsymp}
\hat G_\rm{prim}(x)
	= 
	\left\{
	\begin{aligned}
	&\hat g_{\frak 3} \, x^{\omega_{\frak 3}} \big[ 1 + O(x) \big] ,
	&\qquad &x \to 0
	\\
	&\hat g_{\frak 4} \, x^{\omega_{\frak 4}} \big[ 1 + O(x) \big] ,
	&\qquad &x \to -n
	\end{aligned}
	\right.
\ee
The exponents $\omega_\frh$, $\frh = {\frak 3,\frak 4}$, correspond to the dimensions $\Delta(X^\frh) = h(X^\frh) + \tilde h(X^\frh)$ of the fields $X^\frh$ in the OPE
\be	\label{Channu0Prim}
\begin{aligned}
{\scr O}_{[2]}(u) \s^\fra_{[n]}(0)
&	= 
	C_{{\scr O}_2, \s^\fra_n, X^{\frak 3}_{n-1}} \ |u|^{\Delta( X^{\frak 3}) - \Delta({\scr O}_2) - \Delta(\s^\fra_n)} \ X^{\frak 3}_{[n - 1]}(0) 
\\
&
	+
	C_{{\scr O}_2, \s^\fra_n, X^{\frak 4}_{n+1}} \ |u|^{\Delta( X^{\frak 4}) - \Delta({\scr O}_2) - \Delta(\s^\fra_n)} \ X^{\frak 4}_{[n + 1]}(0) 
\end{aligned}
\ee 
which can be read by substituting $x_\frh(u)$ from (\ref{Channid3}), 
\be
h(X^{\frak 3}_{[n-1]}) = \frac{\omega_{\frak 3}}{n - 1 } + h({\scr O}_{[2]}) + h(\s^\fra_{[n]}) ,
\qquad
h(X^{\frak 4}_{[n+1]}) = \frac{\omega_{\frak 4}}{n + 1 } + h({\scr O}_{[2]}) + h(\s^\fra_{[n]}) .
\ee
The constants $\hat g_\frh$ in (\ref{hatGprimAsymp}) determine structure constants $ C_{{\scr O}_2, \s^\fra_n, X^\frh_{n\pm1}}$.

The effect of the fractional modes in hte RHS of (\ref{hatGscrDhatGprim1}) is fully encapsulated in
\be	\label{DefoCoeffscD}
{\scr D}^{\{a;k;\fra\}}(x) 
	=
	\sum_{\ell, \ell' = 0}^{k+R_\fra} 
	A^{(\infty)}_{k-\ell'}
	A^{(0)}_{k-\ell}
	B^{\{a;\fra\}}_{\ell',\ell} 
	=
	\sum_{\ell, \ell' = -R_\fra}^{k} 
	A^{(\infty)}_{\ell'}
	A^{(0)}_{\ell}
	B^{\{a;\fra\}}_{k-\ell',k-\ell} .
\ee
Here $B_{\ell',\ell}$ are a set of coefficients for which we have given explicit formulas, for instance in (\ref{Jminsigma}).
We will now examine the fate of ${\scr D}(x)$ in the limits (\ref{xaxp4ns}).

For functions with the interaction operator the $B$-coefficients are proportional to Kronecker deltas, so want to know the behavior of the ``diagonal'' product 
$A^{(\infty)}_r A^{(0)}_r$, given by Eq.(\ref{A0AInfkk}).
It is easy to see that
\be	\label{A0AInfkkxtominn}
\begin{aligned}
&
\lim_{x \to -n} A_{\ell}^{(0)}A_{\ell}^{(\infty)} 
	= 
\\
&
	(x+n)^{- \frac{\ell}{n}(n+1) }
	\left(\frac{-n}{n+1} \right)^{\frac{\ell}{n} (n+1)}
	\frac{ (\ell/n)_{\ell}}{\ell!}
	\sum_{q=1}^{\ell}
	(-)^q
	{\cal C}_{\ell,q}
	(-\ell/n)_{q} 
	\Big[1 + O(x+n) \Big]
\end{aligned}
\ee
(recall ${\cal C}_{\ell,\ell} = \ell!$)
while
\be	\label{A0AInfkkxto0}
\begin{aligned}
&
\lim_{x \to 0} A_{\ell}^{(0)}A_{\ell}^{(\infty)} 
	= 
\\
&
	x^{-\frac{\ell}{n}(n-1) }
	\left(\frac{-n}{n-1}  \right)^{\frac{\ell}{n} (n-1)}
	\frac{(-\ell/n)_{\ell}}{\ell!}
	\sum_{p=1}^{\ell}
	(-)^p
	{\cal C}_{\ell,p}
	(\ell/n)_{p} 
	\Big[1 + O(x) \Big]
\end{aligned}
\ee
For functions with NS chirals, the product of $A$-coefficients is not diagonal, but it is not difficult to check, using Eqs.(\ref{A0Infn22n}), that we have the same powers of $x$ and $x+n$ as above, in the corresponding limits.
For ${\scr D}^{\{a;k\}}(x)$ in (\ref{DefoCoeffscD}), the leading term corresponds to the highest value of $\ell = \ell' = k$,
\be	\label{DcoeffAsymp}
\begin{aligned}
{\scr D}^{\{a;k;\fra\}}(x)
	&=
	\frac{1}{x^{\frac{k}{n}(n-1)}} \Big( C_{\frak 3}^{\{a;k;\fra\}} + O(x) \Big)
&\qquad
&\text{for $x \to 0$}
\\
{\scr D}^{\{a;k;\fra\}}(x)
	&=
	\frac{1}{(x+n)^{\frac{k}{n}(n+1)}} \Big( C_{\frak 4}^{\{a;k;\fra\}} + O(x+n) \Big)
&\qquad
&\text{for $x \to -n$}
\end{aligned}
\ee
where $C_\frh$ are constants.
For instance, Eq.(\ref{J3sigmaINT}) gives
\be	\label{Cfrakra}
\begin{aligned}
C_{\frak 3}^{\{3;k;\fra\}} 
	&= 
	- (q^\fra_n)^2
	\left(\frac{-n}{n-1}  \right)^{\frac{k}{n} (n-1)}
	\frac{(-k/n)_{k}}{k!}
	\sum_{p=1}^{k}
	(-)^p
	(k/n)_{p} \;
	{\cal C}_{k,p} ,
\\
C_{\frak 4}^{\{3;k;\fra\}} 
	&= 
	- (q^\fra_n)^2
	\left(\frac{-n}{n+1}  \right)^{\frac{k}{n} (n+1)}
	\frac{(-k/n)_{k}}{k!}
	\sum_{p=1}^{k}
	(-)^p
	(-k/n)_{p} \;
	{\cal C}_{k,p}
	.
\end{aligned}
\ee

Combined with (\ref{hatGprimAsymp}), Eqs.(\ref{DcoeffAsymp}) determine the dimension of the fields in the OPE (\ref{Channu0}),
\be	\label{Channu0Primkn}
\begin{aligned}
	{\scr O}_{[2]}(u) 
	\lord J^a_{-\frac{k}{n}} \s^\fra_{[n]} \rord (0)
&	= 
	C_{{\scr O}_2, \s^\fra_n, Z^{\frak 3}_{n-1}} \ 
	|u|^{\Delta( Z^{\frak 3}_{n-1}) - \Delta({\scr O}_2) - \Delta(J\s^\fra_n)} \ 
	Z^{\frak 3}_{[n - 1]}(0) 
\\
&
	+
	C_{{\scr O}_2, \s^\fra_n, Z^{\frak 4}_{n+1}} \ 
	|u|^{\Delta( Z^{\frak 4}_{n+1}) - \Delta({\scr O}_2) - \Delta(J\s^\fra_n)} \ 
	Z^{\frak 4}_{[n + 1]}(0) 
\end{aligned}
\ee 
which are
\be
\begin{aligned}
h(Z^{\frak 3}_{[n-1]}) 
&	= \frac{\omega_{\frak 3} - \frac{k}{n} (n - 1)}{n - 1} + h({\scr O}_2) + h(J^a_{-k/n}\s^\fra_{[n]}) ,
\\
h(Z^{\frak 4}_{[n+1]}) 
&	= \frac{\omega_{\frak 4} - \frac{k}{n} (n+1)}{n+1} + h({\scr O}_2) + h(J^a_{-k/n}\s^\fra_{[n]}) ,
\end{aligned}
\ee
Since $h(J^a_{-k/n}\s^\fra_{[n]}) = h^\fra_n + k/n$, we find
\be	\label{hZhXeq}
h(Z^{\frak 3}_{[n-1]}) = h(X^{\frak 3}_{[n-1]}) ,
\qquad\qquad
h(Z^{\frak 4}_{[n+1]})  = h(X^{\frak 4}_{[n+1]}) .
\ee

Let us give an illustration.
It is well-known that the OPE algebra of NS chiral fields forms a ring \cite{Dabholkar:2007ey}. For example, 
\be	\label{OPEsmsm}
\big[ \s^-_{[2]} \big] \times \big[\s^-_{[n]} \big]
	=
	\big[ \s^+_{[n-1]} \big]
	+
	\big[ \s^-_{[n+1]} \big] .
\ee
For the $J^3_{-k/n}$ excitation of $\s^-_{[n]}$ Eq.(\ref{hZhXeq}) gives
\be	\label{Channu0A}
\big[ \s^-_{[2]} \big]
	\times 
	\big[ \lord J^3_{-\frac{k}{n}} \s^-_{[n]}\rord  \big]
	= 
	[Z^{{\frak 3}}_{[n-1]} ]
	+
	[Z^{\frak 4}_{[n+1]} ] 
\quad \text{where} \quad
	\begin{aligned}[t]
	h(Z^{\frak 3}_{[n-1]}) &= \frac{n}{2} ,
	\\
	h(Z^{\frak 4}_{[n+1]}) &= \frac{n}{2} .
	\end{aligned}	
\ee 
The total charge in the LHS is $q^-_2 + q^-_n = \frac12 + \frac12(n-1) = \frac12 n$, which must also be the charges of the fields in the RHS.
In this case the $Z$ fields have equal dimensions and R-charges, i.e.~they are, again, chiral primaries:
\be
Z^{\frak 3}_{[n-1]} =  \s^+_{[n-1]},
\qquad
Z^{\frak 4}_{[n+1]} = \s^-_{[n+1]} .
\ee
This might seem counterintuitive at first sight, but it is just a confirmation of the fusion rules (\ref{OPEsigmmInt}) predicted by \cite{deBeer:2019ioe}.
That is, the RHS of the fusion rule (\ref{OPEsmsm}) is actually incomplete, the fields $J^3_{-k/n} \s^-_{[n]}$  are missing. Note that, as discussed in \S\ref{SectOnBeingPrimary}, these fields are actually primary. The structure constants 
\be	\label{StrucFina}
\langle \s^{\mp \dagger}_{[n\pm1]}(\infty) \ \s^-_{[n]}(1) \ \lord J^3_{-k/n} \s^-_{[n]} \rord (0) \rangle
\ee
can be found from the asymptotics of ${\scr D}$ as in (\ref{DcoeffAsymp}).
In \cite{deBeer:2019ioe} the authors have computed, for several values of $k$, a series of structure constants 
$\langle \s^{\mp \dagger}_{[n\pm1]} \ \s^-_{[n]} \ {\cal O}_{[n+1]} \rangle$
where ${\cal O}_{[n+1]}$ is a specific combination of $J^3_{-k/n}$ and $L_{-k/n}$ obtained by spectral flow of chiral primaries. If we isolate the contributions from $J^3$ from those of $L$, we expect that their results should match the results for (\ref{StrucFina}) that we find here, but we have not checked this explicitly.  There is, however, one simple check that we can do.  One of the notable facts found in  \cite{deBeer:2019ioe} is the absence of the field $J^3_{-1/n} \s^-_{[n]}$ from the RHS of the fusion (\ref{OPEsigmmInt}).
For $k = 1$, we have displayed the function ${\scr D}^{\{3,1\}}_\NS(x)$ explicitly in (\ref{scrD31NS}). 
For $\fra = \frb = -$, it reads
\be
\begin{aligned}
{\scr D}^{\{3;1\}}_\NS(x)
&	= 
	- \left[
	\frac{x ( x+n-1)}{(x+n)(x-1)} 
	\right]^{\frac{1}{n}}
	\Bigg[
	\frac12
	+
	\frac{\big( n-1 - (2x+n-1) \big)^2}
		{4 x (x-1)(x+n)(x+n-1)}
	\Bigg] 
\end{aligned}
\ee 
The expansions around $x = 0$ and $x = -n$ are 
\be
\begin{aligned}
{\scr D}^{\{3;1\}}_\NS(x)
&	= 
	x^{\frac{n+1}{n}} 
	\left[
	(1-n)^{-\frac{3n-1}{n}} n^{\frac{n-1}{n}}
	+ O(x)
	\right]
\\
{\scr D}^{\{3;1\}}_\NS(x)
&	= 
	(x+n)^{\frac{n-1}{n}} 
	\left[
	\frac{(-)^{\frac{n-1}{n}} n^{\frac{n+1}{n}} }{ (n+1)^{\frac{3n+1}{n}}}
	+ O(x+n)
	\right]
\end{aligned}
\ee 
They match (\ref{DcoeffAsymp}), with both $C^{\{3;1;-\}}_\frh = 0$. Indeed, here both expansions in parenthesis in (\ref{DcoeffAsymp}) begins at second order, for example we can rewrite the above as
\be
{\scr D}^{\{3;1\}}_\NS(x)
	=
	\frac{1}{x^{\frac{n-1}{n}} }
	\Big(
	(1-n)^{-\frac{3n-1}{n}} n^{\frac{n-1}{n}} \ x^2
	+ O(x^3)
	\Big)
\ee
The vanishing of $C^{\{3;1;-\}}_\frh = 0$ implies the vanishing of (\ref{StrucFina}), hence the absence of $J^3_{-1/n} \s^\pm_{[n\mp1]}$ from the fusion (\ref{OPEsigmmInt}), in accordance with \cite{deBeer:2019ioe}.

\section{Ramond fields}

The fact that NS chirals and anti-chirals do not lift to Kac-Moody primaries raises the question of which states do. The answer is Ramond ground states.

\bigskip

The Ramond ground states in the $n$-twisted sector are a set of four states $\ket{R^\fra}_n$ with degenerate dimension $h^\R_n = \frac14 n$, and R-charges
\be
q^{\R \pm}_n = \pm \tfrac{1}{2},
\qquad 
q^{\R \dot A}_n = 0.
\ee
(There are analogous states in the anti-holomorphic sector.)
The corresponding fields will be denoted by
$R^{\fra}_{(n)} \cong \ket{R^\fra}_n$. They can be written in bosonized form as
\bsub\label{noninRamond}
\begin{align}
R^{\pm}_{(n)} (z) 
	&=
	e^{
	 \pm \frac{i}{2n} \sum_{I = 1}^n
	 ( \phi_{1,I} - \phi_{2,I} )
	} 
	\s_{(n)}(z) ,
\\
R^{\dot 1}_{(n)}(z)
	&=
	e^{
	 - \frac{i}{2n} \sum_{I = 1}^n
	 \left( \phi_{1,I} + \phi_{2,I} \right)
	} 
	\s_{(n)}(z) ,
\\
R^{\dot 2}_{(n)}(z)
	&=
	e^{
	 + \frac{i}{2n} \sum_{I = 1}^n
	 \left( \phi_{1,I} + \phi_{2,I} \right)
	} 
	\s_{(n)}(z) .
\end{align}
\esub
The ${\cal N}=4$ superalgebra has a spectral flow automorphism that acts on the R-symmetry and Virasoro modes as \cite{Schwimmer:1986mf}
\be
\begin{aligned}
J^3_\ell &\mapsto J^3_\ell - \tfrac1{12} c \xi \delta_{\ell,0} 
\\
J^\pm_\ell &\mapsto J^\pm_{\ell \mp \eta}
\\
L_\ell &\mapsto L_\ell - \tfrac1{24} c \xi^2 \delta_{\ell,0} 
\end{aligned}
\ee
The Ramond fields are related to the NS chiral fields (\ref{NSchiralss}) by a spectral flow with $\xi=+1$, or to NS anti-chirals by a flow with $\xi=-1$.
This allows us to obtain correlation functions with Ramond fields by applying the appropriate spectral flow to a function with NS chirals/anti-chirals.

If we want to explicitly compute functions with $n$-twisted Ramond ground states in the same way we did above for NS chirals, we have to take into account a small subtlety \cite{Lima:2021wrz}.
When we define a state/field on the $n$-twisted sector, we are implicitly ignoring a factor of $(N-n)$ \emph{untwisted} states/fields corresponding to the other copies in the orbifold that do not enter the cycle $(n)\in S_N$. 
The NS chiral states $\ket{\s^\fra}_n$ defined in (\ref{NSchiralStates}), which belong to the Hilbert space sector corresponding to the permutation $(n) = (1,2,\cdots, n)$,
should be understood, implicitly, as the tensor product
\be	\label{ketsfravarnoth}
\ket{\s^\fra}_n 
	= \ket{\s^\fra}_n \otimes (\ket{\s^-}_1)^{N-n}
	= \ket{\s^\fra}_n \otimes (\ket{\varnothing})^{N-n} ,
\ee
where the untwisted state $\ket{\s^-}_1$ is the NS chiral whose dimension, according to Eq.(\ref{NSchiralStates}), is $h^-_1 = 0$.
That is, the vacuum, $\ket{\varnothing}$, belongs to the NS sector. In contrast, in the Ramond sector the implicit untwisted fields should not be taken as the vacuum; instead, they should be one of the untwisted Ramond ground states with dimension $h^\R_1 = \frac14$, for example, the untwisted state $\ket{R^-}_1$, with charge $h^{\R-}_1 = - \frac12$, which is the image of the vacuum $\ket{\s^-}_1$ under spectral flow.
Thus the spectral flowed image of (\ref{ketsfravarnoth}) is the $n$-twisted Ramond state
\be	\label{Rframin}
\ket{R^\fra}_n \otimes (\ket{R^-}_1)^{N-n} .
\ee

Now, the subtlety is that the \emph{untwisted copies do not fully factorize from correlation functions,} because of the fusion with the other twisted fields in the correlator. 
Consider, for example, the following four-point function with excitations of $n$-twisted Ramond ground states:
\be	\label{GRketpspspsZ}
G_\R(u) = 
	\big\langle {\scr R}_{k;n}^{a; \fra} \big| 
	\; {\scr O}^1_{[2]}(1)
	\; {\scr O}^2_{[2]}(u) \; 
	\big| {\scr R}_{k;n}^{a; \fra} \big\rangle ,
\quad
\big| {\scr R}_{k;n}^{a; \fra} \big\rangle
	\equiv
	\Big( J^a_{-\frac{k}{n} } \ket{R^\fra}_n \Big) \otimes (\ket{R^-}_1)^{N-n} .
\ee
It lifts to a combination of \emph{six}-point functions
\be	\label{GtsumellRB}
\hat G^\R_{\ell',\ell}({\bs t})
	=
	\Big\langle
	 J^{a\dagger}_{\ell'} \hat R^{\fra \dagger}_{n} (\infty) 
	\hat R^{+}_{1} (t_\infty)
	\hat {\scr O}^1_{2}(t_1) 
	\hat {\scr O}^2_{2}(t_2) 
	\hat R^{-}_1 (t_0)
	J^a_{- \ell} \hat R^{\fra}_n (0)
	\Big\rangle .
\ee
The points $t_0, t_\infty$ are such that $z(t_0) = 0$ and $z(t_\infty) = \infty$, as displayed in (\ref{Mappingttinfo}), but here these are \emph{unramified} points on the covering defined by our ``single-cyle map''  (\ref{coverm}).
Thus, in a sense, the functions (\ref{GRketpspspsZ}) are more similar to double-cycle functions, but with the single-cycle map. 
This lifting to a more-point function is, of course, generic for any covering map, as the permutations will generically mix the untwisted copies in (\ref{Rframin}).

The Ramond fields (\ref{noninRamond}) lift to
\bsub \label{0022RamchiCOVER}
\begin{align}
\hat R^{\pm}_{n} (t) 
	&=
	b_*^{- \frac{1}{4 n}}
	e^{ \pm \frac{i}{2} (\phi_{1} - \phi_{2} )} (t)
\\
\hat R^{\dot 1}_{n}(t)
	&=
	b_*^{- \frac{1}{4 n}}
	e^{- \frac{i}{2}  (\phi_{1} + \phi_{2}) } (t)
\\
\hat R^{\dot 2}_{n}(t)
	&=
	b_*^{- \frac{1}{4 n}}
	e^{+ \frac{i}{2}  (\phi_{1} + \phi_{2}) } (t)
\end{align}
\esub
It is easy to check that, unlike the NS chirals, these lifted Ramond fields are, in fact, Kac-Moody primaries:
\be	\label{OPEJ3VRam}
J^a(t) \hat R^{\fra}_n(t_*) 
	= \frac{ \sum_{\frc} [{\scr J}^a]^\fra{}_{\frc} \hat R^{\frc}_n(t_*)   }{t-t_*} + \text{Regular}
\ee
For the doublet $\hat R^\a_n$ (i.e.~$\fra = \a = \pm$) the representation matrices
$[{\scr J}^a]^\a{}_{\ga} = \frac12 [\s^a]^\a{}_\ga$
are Pauli matrices,
and for the singlets $\hat R^{\dot A}_n$ the representation is trivial
$[{\scr J}^a]^\a{}_{\ga} = 0$.
In other words,
\be	\label{J0JnViplusRam}
\begin{aligned}
	\lord  J^-_{0} \hat R^+_n\rord
	&= \hat R^-_n ,
	&\qquad
	\lord  J^+_{0} \hat R^-_n\rord
	&= \hat R^+_n ,
	&\qquad
	\lord  J^3_{0} \hat R^\fra_n\rord 
	&= q^{\R \fra}_n \hat R^\fra_n ,
	&
	&
\\
	\lord  J^+_{\ell} \hat R^+_n\rord  
	&= 0 ,
	&\qquad
	\lord  J^-_{\ell} \hat R^-_n\rord
	&= 0 ,
	&\qquad
	\lord  J^3_{\ell} \hat R^\fra_n\rord
	&= 0 ,
	&\quad
	\text{for\ } \ell &> 0
\\
	\lord  J^\pm_{\ell'} \hat R^{\dot A}_n\rord 
	&= 0 ,
	&
	&
	&
	&
	&\quad
	\text{for\ } \ell' &\geq 0
\end{aligned}
\ee

It follows that (\ref{GRketpspspsZ}) not only lifts to a finite combination of correlators, but it has the form
\be	\label{GtsumellR}
\hat G_\R({\bs t}) 
	= 
	\sum_{\ell = 0}^{k} 
	\sum_{\ell' = 0}^{k}
	A^{(\infty)}_{k'-\ell'}
	A^{(0)}_{k-\ell}
	\hat G^\R_{\ell',\ell}({\bs t})
\ee
i.e.~the indices $\ell,\ell'$ in (\ref{GtsumellRB}) dot not take negative values.

It also follows that functions containing only Ramond fields and their excitations can always be factorized. More precisely, they can be expressed in terms of correlators involving Ramond fields without any excitation, but which are ``rotated'' by the ${\scr J}^a$ matrices.
Let us go back to an arbitrary twist configuration (provided the covering is a sphere), and also take the most general configuration of $\frak{su}(2)$ indices possible:
\be	\label{GRamonRRRra}
G^{\{a; \fra_i; \frb_i\}}_\R(u) = 
	\big\langle {\scr R}_{k;n'}^{a; \fra'} \big| 
	\; R^{\frb_1}_{[n_1]}(1)
	\; R^{\frb_2}_{[n_2]}(u) \; 
	\big| {\scr R}_{k;n}^{a; \fra} \big\rangle ,
\ee
This lifts to a combination of 
\be
\hat G^\R_{\ell',\ell}({\bs t})
	=
	\Big\langle
	\lord J^{a \dagger}_{\ell'} \hat R^{\fra'}_{n'} \rord (\infty) 
	\hat R^{+}_{1} (t_\infty)
	\hat R^{\frb_1}_{2}(t_1) 
	\hat R^{\frb_2}_{2}(t_2) 
	\hat R^{-}_1 (t_0)
	\lord J^a_{- \ell} \hat R^{\fra}_n \rord (0)
	\Big\rangle .
\ee
Using Ward identities along with (\ref{J0JnViplusRam}), 
\be
\begin{aligned}
\hat G^\R_{\ell',\ell}({\bs t})
&	=
	-
	\Big\langle
	\hat R^{\fra'}_{n'} (\infty) 
	\hat R^{+}_{1} (t_\infty)
	\hat R^{\frb_1}_{2}(t_1) 
	\hat R^{\frb_2}_{2}(t_2) 
	\hat R^{-}_1 (t_0)
	\lord J^{a'}_{\ell'} \lord J^a_{- \ell} \hat R^{\fra}_n \rord\rord (0)
	\Big\rangle
\\
&\quad
	- 
	t_\infty^{\ell'}
	[ {\scr J}^{a\dagger}]^+{}_\b
	\Big\langle
	\hat R^{\fra'}_{n'} (\infty) 
	\hat R^{\b}_{1} (t_\infty)
	\hat R^{\frb_1}_{2}(t_1) 
	\hat R^{\frb_2}_{2}(t_2) 
	\hat R^{-}_1 (t_0)
	\lord J^a_{- \ell} \hat R^{\fra}_n \rord (0)
	\Big\rangle
\\
&\quad
	- 
	t_1^{\ell'}
	[ {\scr J}^{a\dagger}]^{\frb_1}{}_\frc
	\Big\langle
	\hat R^{\fra'}_{n'} (\infty) 
	\hat R^{+}_{1} (t_\infty)
	\hat R^{\frc}_{2}(t_1) 
	\hat R^{\frb_2}_{2}(t_2) 
	\hat R^{-}_1 (t_0)
	\lord J^a_{- \ell} \hat R^{\fra}_n \rord (0)
	\Big\rangle
\\
&\quad
	- 
	t_2^{\ell'}
	[ {\scr J}^{a\dagger}]^{\frb_2}{}_\frc
	\Big\langle
	\hat R^{\fra'}_{n'} (\infty) 
	\hat R^{+}_{1} (t_\infty)
	\hat R^{\frb_1}_{2}(t_1) 
	\hat R^{\frc}_{2}(t_2) 
	\hat R^{-}_1 (t_0)
	\lord J^a_{- \ell} \hat R^{\fra}_n \rord (0)
	\Big\rangle
\\
&\quad
	- 
	t_0^{\ell'}
	[ {\scr J}^{a\dagger}]^{-}{}_\b
	\Big\langle
	\hat R^{\fra'}_{n'} (\infty) 
	\hat R^{+}_{1} (t_\infty)
	\hat R^{\frb_1}_{2}(t_1) 
	\hat R^{\frb_2}_{2}(t_2) 
	\hat R^{\b}_1 (t_0)
	\lord J^a_{- \ell} \hat R^{\fra}_n \rord (0)
	\Big\rangle
\end{aligned}
\ee
Here summation over repeated indices in the matrix products is implied.
The function in the first line can be reduced by commuting 
$[J^{a\dagger}_{\ell'} , J^a_{-\ell}]$,
which, together with (\ref{J0JnViplusRam}), results in a field proportional to $\hat R^{\fra}_n (0)$.
In the next four lines, we move the mode $J^a_{-\ell}$ to the other fields using Ward identities, and reduce the correlators to a combination of correlators without excitations, multiplied by representation matrices. Each line produces five more terms, with an obvious structure.
Thus we conclude that (\ref{GRamonRRRra}) lifts to
\be	\label{GtsumellRBscrD}
\begin{aligned}
\hat G^{\{a; \fra, \fra'; \frb_1, \frb_2\}}_\R({\bs t}) 
&	= 
	\sum_{\ell = 0}^{k} 
	\sum_{\ell' = 0}^{k}
	A^{(\infty)}_{k'-\ell'}
	A^{(0)}_{k-\ell}
	\hat G^\R_{\ell',\ell}({\bs t})
\\
	&=
	\big[ {\scr D}^a_\R({\bs t}) \big]^{\fra, \fra'; \frb_1, \frb_2}_{\frr, \a, \frs, \frt, \b, \frp}
	\Big\langle
	\hat R^{\frr}_{n'} (\infty) 
	\hat R^{\a}_{1} (t_\infty)
	\hat R^{\frs}_{2}(t_1) 
	\hat R^{\frt}_{2}(t_2) 
	\hat R^{\b}_1 (t_0)
	\hat R^{\frp}_n (0)
	\Big\rangle
\end{aligned}
\ee
where ${\scr D}^a_\R({\bs t})$ here should be understood as a linear operator, and summation over its lower indices is implied.

Of course, functions involving Ramond fields and the deformation operator can also be reduced in exactly the same way. They are actually slightly simpler since, as we have seen, the action of the representation matrices in $\Oincov$ is trivial.

\bigskip

In conclusion, the Ramond fields are, in a sense, more natural objects on the covering surface with regard to the action of the R-symmetry currents, even though the twist fusions make correlators with Ramond states more complicated, proliferating the number of fields that must be taken into account (in our example, a four-point function becomes a six-point function).
When there are no excitations, it is usually very efficient to compute functions in the NS sector, and then spectrally flow the result to the Ramond sector if needed but here it is, arguably, more natural to do the opposite: in the presence of R-symmetry excitations one computes the functions in the R-sector, where the result is simpler, and then one flows back.

\bigskip

\noindent
{\bf Acknowledgments}
\\
The work of the authors is partially supported by the Bulgarian NSF grant  KP-06-H88/1.
This study was financed in part by the Coordenação de Aperfeiçoamento de Pessoal de Nível Superior -- Brasil (CAPES) -- Finance Code 001.

\bigskip

\appendix 

\section{Notations for the D1-D5 CFT} 	\label{SectNotationsD1D5}

We use the same conventions for the ${\cal N} = (4,4)$ SCFT as in \cite{Avery:2010qw}.
The $\cal N = (4,4)$ SCFT can be realized in terms of a collection of four real free bosons, four real free holomorphic fermions, and four real free anti-holomorphic fermions.
The central charge is
$c = 6 .$
The theory has R-symmetry group
$SU(2)_L \times SU(2)_R$ and an automorphism group 
$SO(4)_I = SU(2)_1 \times SU(2)_2$.
The bosons can be grouped according to their transformation under $SU(2)_1 \times SU(2)_2$ as
$X_{\dot A  A}$, and the fermions can be grouped by R-symmetry into a pair of complex spinors $\psi^{\a \dot A}(z)$, with analogous constructions for the right-moving sector for $\tilde \psi^{\dot\a \dot A}(\bar z)$.
That is, indices $\a = + , -$ and $\dot \a = \dot +, \dot -$ transform as a doublets of SU(2)$_L$ and SU(2)$_R$,  and $A=1,2$ and $\dot A=\dot1,\dot2$ transform as doublets of SU(2)$_1$ and SU(2)$_2$, respectively.
As usual \cite{Lunin:2001pw}, we will write the fermions in terms of a pair of chiral bosons%
	\footnote{%
	A small difference in notation that we adopt in comparison with \cite{Lunin:2001pw, Avery:2010qw} is that we call $\phi_1, \phi_2$ the fields they call $\phi_5,\phi_6$, respectively.}
$\phi_{1}(z)$ and $\phi_{2}(z)$ normalized such that
\be
\big\langle \phi_r(z) \phi_s(z') \big\rangle
	= - \delta_{rs} \log(z-z') .
\ee
In the free CFT, the fermions are given by the exponentials (always implicitly normal-ordered)
\be	\label{FermionsBoson}
\psi^{+ \dot 1}(z) = e^{-i \phi_{2}(z)} , 
\quad
\psi^{+ \dot 2}(z) = e^{i \phi_{1}(z)} ,	
\quad
\psi^{- \dot 1}(z) = e^{- i \phi_{1} (z)} , 
\quad
\psi^{- \dot 2}(z) = - e^{i\phi_{2} (z)} .
\ee
We are ignoring cocycles, see \cite{Burrington:2012yq, Burrington:2015mfa}.
The conserved currents are expressed in terms of the basic fields as
\bsub\label{TJGasXpsi}
\begin{align}
T(z) &= \tfrac{1}{2} \e_{\dot A \dot B} \e_{AB} \pa X^{\dot A A} \pa X^{\dot B B} + \tfrac{1}{2} \e_{\dot A \dot B} \e_{\a\b} \psi^{\a \dot A} \pa \psi^{\b \dot B}  \label{TJGasXpsiT}
\\
J^a(z) &= \tfrac{1}{4} \e_{\dot A \dot B} \e_{\a\b} \psi^{\a \dot A} [\s^{*a}]^\b{}_\ga \psi^{\ga \dot B}
\\
G^{\a A}(z) &= \e_{\dot A \dot B} \psi^{\a \dot A} \pa X^{\dot B A}
	\label{GaAz}
\end{align}\esub
with similar expressions for the antiholomorphic sector.

The symmetric orbifold is made by taking $N$ independent copies $I= 1, \cdots, N$ of the ${\cal N} = (4,4)$ SCFT, then called a `seed CFT', and identifying copies under the action of the symmetric group $S_N$. This amounts to imposing non-trivial monodromies associated with permutations $g \in S_N$, implemented by the insertion of `bare' twist fields $\s_g(z)$, such that, for any given field ${\scr O}_I$ in the $I$th copy of the CFT,
\be
{\scr O}_I(e^{2\pi i} z) \s_g(0) = {\scr O}_{g(I)}(z) \s_g(0) .
\ee
From the Hilbert space point of view, the `bare twists' $\s_g$ correspond to ground states  of twisted sectors. 
A fundamental role is played by twists corresponding to the single-cycle permutations, because any $g \in S_N$ can be uniquely decomposed into disjoint cycles. We will denote the twists corresponding to the cycles $g = (1,\cdots, n)$ by
$\s_{(n)}$.
They can be shown to have conformal weights
\be\label{twistdim}
h_n^\s = \frac{1}{4} \Big( n - \frac{1}{n} \Big) = \tilde h^\s_n ,
\ee
We will say that the sector of Hilbert space with ground state $\s_{(n)}$ is the `$n$-twisted sector'. Since only $n$ copies effectively contribute, we can say that the central charge in this sector is $c = 6n$.

\section{Two Ward identities}	\label{AppWardId}

Assume the covering surface has the topology of a sphere. 
Consider the insertion of operators at points $t_i$ on the covering surface, with $i = 0,1,\cdots, Q$ and $t_0 = 0$, plus an excited operator at $t = \infty$,
\be
\begin{aligned}
\Big\langle 
\lord  J^a_{\ell} \hat {\scr O}_\infty \rord (\infty)
\prod_{i=0}^Q \hat {\scr O}_i(t_i) 
\Big\rangle 
&=
\oint_\infty \frac{dt}{2\pi i} \, t^\ell
\Big\langle 
J^a(t) \;  \hat {\scr O}_\infty (\infty) \prod_{i=0}^Q \hat {\scr O}_i(t_i) 
\Big\rangle 
\\
&=
-
\sum_{i =0}^Q
\oint_{t_i} \frac{dt}{2\pi i} \, t^\ell
\Big\langle 
J^a(t)  \hat {\scr O}_\infty (\infty) \hat {\scr O}_Q(t_Q) \cdots \hat {\scr O}_1(t_1) \hat {\scr O}_0(0) 
\Big\rangle .
\end{aligned}
\ee
In the second line, we have pulled the contour around $\infty$ until it splits into contours around the other poles of the integrand, i.e.~around the insertions of the fields $\hat {\scr O}_i(t_i)$.
Using (\ref{OPEordering}), we can compute the residues to find
\be	\label{JpWardId}
\begin{aligned}
\Big\langle 
\lord  J^a_{\ell} \hat {\scr O}_\infty \rord (\infty) \prod_{i=0}^Q \hat {\scr O}_i(t_i)
\Big\rangle 
&=
-
\Big\langle 
{\scr O}_\infty (\infty) \;  \hat {\scr O}_Q(t_Q) \cdots \hat {\scr O}_1(t_1) 
	\lord  J^a_\ell \hat {\scr O}_0 \rord (0) 
\Big\rangle 
\\
&
-
\sum_{r = 0}^\ell {\ell \choose r}
\Bigg[
t_Q^{\ell-r} 
\Big\langle 
\hat {\scr O}_\infty (\infty) \; \lord  J^a_r \hat {\scr O}_Q\rord (t_Q)  \cdots \hat {\scr O}_0(0) 
\Big\rangle 
+ \cdots
\\
&\qquad\qquad
+
t_1^{\ell-r} 
\Big\langle 
\hat {\scr O}_\infty (\infty) \; \hat {\scr O}_Q(t_Q) \cdots \lord  J^a_r \hat {\scr O}_1\rord (t_1)  \hat {\scr O}_0(0)
\Big\rangle 
\Bigg]
\\
\end{aligned}
\ee
By a similar reasoning, if we have a descendant at a finite point $t_0$, we can pull the contours and use the binomial expansion to find
\be	\label{JpWardIdfin}
\begin{aligned}
\Big\langle 
\hat  {\scr O}_\infty(\infty) \hat {\scr O}_Q(t_Q) \cdots \hat {\scr O}_1(t_1) & \, \lord J_{-\ell} \hat  {\scr O}_0\rord (t_0)  
\Big\rangle 
\\
	=
	-
	\sum_{r \geq 0}
	(-)^{r}
	{\ell + r -1 \choose r}
&	\Bigg[	
	(t_1 - t_0)^{r-\ell}
	\Big\langle
	\hat  {\scr O}_\infty(\infty) 
	 \hat {\scr O}_Q(t_Q) \cdots 
	\lord J^a_r \hat {\scr O}_1\rord (t_1) 
	\hat  {\scr O}_0(t_0)
	\Big\rangle
+ \cdots 
\\
& +
	(t_Q - t_0)^{r-\ell}
	\Big\langle
	\hat {\scr O}_\infty(\infty) 
	\lord J^a_r \hat {\scr O}_Q\rord (t_Q)  \cdots 
	\hat {\scr O}_1(t_1) 
	\hat {\scr O}_0(t_0)
	\Big\rangle
\Bigg]
\\
& -
	\Big\langle
	\lord J^a_{-\ell} \hat {\scr O}_\infty\rord (\infty)
	\hat {\scr O}_Q(t_Q) \cdots 
	\hat {\scr O}_1(t_1) 
	\hat {\scr O}_0(t_0)
	\Big\rangle
\end{aligned}
\ee

We should emphasize that the covering surface must have the topology of a sphere, so that there are no topological obstructions to the contour deformations. 
Generic $Q$-point functions with twisted fields containing a total number $R$ of cycles will have contributions from covering surfaces of different genera.
In the large-$N$ limit, the contribution from surfaces of genus zero. In other words, our contour manipulations above work at leading order in the large $N$ limit.

\section{Bell polynomials and Faà di Bruno's formula}	\label{AppBellPoly}

The \emph{`incomplete exponential Bell polynomials'} are defined by \cite{charalambides2018enumerative}
\be	\label{BellPoluincomdef}
\begin{aligned}
\Bell_{\ell, m} \big[ x_1, \cdots, x_{\ell-m+1} \big]
	&= 
	\sum_{[j]}
	\prod_{n=1}^{\ell-m+1}
	\frac{\ell!}{(n!)^{j_n} j_n!} \, x_n^{j_n}
\\
	&= 
	\sum_{[j]}
	\frac{\ell!}{j_1! j_2 ! \cdots j_{\ell-m+1}!}
	\left( \frac{x_1}{1!} \right)^{j_1}
	\left( \frac{x_2}{2!} \right)^{j_2}
	\cdots
	\left( \frac{x_{\ell-m+1}}{(\ell-m+1)!} \right)^{j_{\ell-m+1}}
\end{aligned}
\ee
where the sum $\sum_{[j]}$ is over all $j_n \geq 0$, restricted by the following pair of conditions
\be	\label{BellPoluincomdefCond}
\begin{aligned}
j_1 + \cdots j_{\ell - m +1} &= m ,
\\ 
j_1 + 2 j_2 + 3 j_3 + \cdots + (\ell-m+1) j_{\ell-m+1} &= \ell .
\end{aligned}
\ee
For economy of notation, one may define
$\Bell_{\ell, m}  = \Bell_{\ell, m} \big[ x_1, \cdots, x_{\ell} \big]$,
and extended the product in the RHS of (\ref{BellPoluincomdef}) from $n =1$ to $n = m$ (in fact, this is how the polynomials are defined in \cite{charalambides2018enumerative}). Since there are no solutions to the equations (\ref{BellPoluincomdefCond}) for $x_a$ with $a > \ell - m +1$, both definitions are equivalent.
The following identity follows from the definition:
\be	\label{IdentBellHomo}
\Bell_{\ell,m}\big[ab x_1, a^2 b x_2 , \cdots, a^{\ell-m+1} b x_{\ell-m+1} \big]
	= a^\ell b^m \Bell_{\ell,m} \big[ x_1, \cdots, x_{\ell-m+1} \big] .
\ee

The polynomials encode a combinatorics information:
the coefficient of the term $x_{a_1} \cdots x_{a_r}$ in $\Bell_{\ell, m}$ is the number of ways of partitioning a set $\{A_1, \cdots, A_\ell \}$ with $\ell$ elements into disjoint unions of $m$ sets with $a_1, a_2, \cdots a_r$ elements each.
A list of the first incomplete exponential Bell polynomials is
{\small
\be
\begin{aligned}
&& \Bell_{1,1} &= x_1, &	&	&&
\\
&& \Bell_{2,1} &= x_2, &\quad \Bell_{2,2} &= x_1^2,	&&
\\
&& \Bell_{3,1} &= x_3, &\quad \Bell_{3,2} &= 3 x_1 x_2,	
	&\quad
	\Bell_{3,3} &= x_1^3 ,
\\
&& \Bell_{4,1} &= x_4, 
	&\quad \Bell_{4,2} &= 4 x_1 x_3 + 3 x_2^2,	
	&\quad \Bell_{4,3} &= 6 x_1^2 x_2 ,
\\
&& &
	&\quad &
	&\quad \Bell_{4,4} &= x_1^4 ,
\\
&& \Bell_{5,1} &= x_5, 
	&\quad \Bell_{5,2} &= 5 x_1 x_4 + 10 x_2 x_3,	
	&\quad \Bell_{5,3} &= 10 x_1^2 x_3 + 15 x_1 x_2^3 ,
\\
&& &
	&\quad &
	&\quad \Bell_{5,4} &= 10 x_1^3 x_2 ,
\\
&& &
	&\quad &
	&\quad \Bell_{5,5} &= x_1^5 ,
\\
\end{aligned}
\ee}%

The Bell polynomials appear notably in the version due to Riordan \cite{riordan1946derivatives} of Faà di Bruno's formula: given two functions $f$ and $g$ and their composition $f(g(t))$, assuming all the necessary derivatives are well-defined we have
\be	\label{FaadiBruno}
\frac{d^\ell}{dt^\ell} f(g(t))
	= 
	\sum_{m=0}^\ell 
	f^{(m)}(g(t))
	\,
	\Bell_{\ell,m} \big[ g'(t) , g''(t) , \cdots , g^{(\ell-m+1)}(t) \big] .
\ee
Here $f'(t) = df/dt$, $f^{(m)}(x) = d^m f / dx^m$, etc.

\bibliographystyle{utphys}
\bibliography{V40_Paper_Fractional_Descendants_Ref} 
\end{document}